\documentstyle[12pt]{article}
\begin{document}
\begin{titlepage}
\begin{centering}
\vspace{2cm}
{\Large\bf Polarisation-Asymmetry Correlation}\\
\vspace{0.5cm}
{\Large\bf in Allowed $\beta$-Decay:}\\
\vspace{0.5cm}
{\Large\bf a Probe for Right-Handed Currents}\\
\vspace{4cm}
J. Govaerts$^{\dag}$, M. Kokkoris$^{\ddag}$ and J. Deutsch$^{\dag}$\\
\vspace{0.5cm}
$^{\dag}${\em Institut de Physique Nucl\'eaire}\\
{\em Universit\'e Catholique de Louvain}\\
{\em B-1348 Louvain-la-Neuve, Belgium}\\
\vspace{0.5cm}
$^{\ddag}${\em Institute of Nuclear Physics}\\
{\em National Center for Scientific Research ``Demokritos"}\\
{\em Attiki, Greece}\\
\vspace{3cm}
\begin{abstract}

\noindent The sensitivity of polarisation-asymmetry correlation experiments
to charged currents of right-handed chirality contributing to allowed
$\beta$-decay is considered in the most general context possible,
independently of any type of approximation
nor of any specific model for physics beyond the Standard
Model of the electroweak interactions.
Results are then particularised
to general Left-Right Symmetric Models, and experimental prospects
offered by mirror nuclei are assessed explicitly
on general grounds. In order of decreasing interest,
the cases of $^{17}$F, $^{41}$Sc and $^{25}$Al are the most
attractive,
providing sensitivities better or comparable to allowed pure
Gamow-Teller transitions, with the advantage however,
that recoil order corrections are smaller in the case of
super-allowed decays.

\end{abstract}

\vspace{30pt}

\end{centering}
\end{titlepage}

\section{Introduction}
\label{Sect1}

\vspace{10pt}

Even though the Standard Model (SM) remains unchallenged by
an impressive body of precision electroweak measurements,
over the years the general consensus has been
that there must exist new physics lurking
behind the horizon of the related problems of the origin of mass
and the chirality structure of the electroweak interactions.
Any embodiment of such new physics would manifest itself trough
deviations from the predictions of the SM for specific
observables and production mechanims, be it at low, intermediate
or high energies.

In the case of semi-leptonic weak interactions at low energies,
it is essentially only the radioactive nucleus which
is available as a laboratory for the search of new physics
through $\beta$-decay\cite{DQ}.
In particular, it has recently been emphasized\cite{Quin} that when
compared to asymmetry measurements,
the {\em relative\/}
longitudinal polarisation of $\beta$ particles emitted
in directions parallel
or antiparallel to the polarisation vector of an oriented nucleus
presents an enhanced sensitivity to a hypothetical
right-handed charged current contribution.
Indeed, one among
other attractive extensions of the SM is
obtained by enlarging the electroweak sector of the latter model
to the gauge group
$SU(2)_L\times SU(2)_R\times U(1)_{B-L}$
in the context of so-called
Left-Right Symmetric Models (LRSM)\footnote{For reviews and references
to the original literature, see for example Refs.\cite{LRSM,Herczeg,Lang},}.
In such a case, the new
physics to be found beyond the SM stems from the existence of additional
charged and neutral gauge $W'$ and $Z'$
and physical Higgs bosons, as well as a host of additional
phenomenological parameters leading to massive neutrinos
and the ensuing leptonic flavour mixing, to new sources of CP violation
o\-ri\-gi\-na\-ting in the Higgs as well as in the Yukawa sectors,
and to other physical effects of interest (see for example
Refs.\cite{Herczeg,Lang}). In spite of their great appeal,
such extensions do not provide an understanding for the origin of mass
nor for the chirality structure of the fundamental electroweak
interactions---now enlarged to include right-handed ones---, even though
parity invariance may {\em effectively\/} be restored in
certain classes of such models for
processes characterised by momenta transfers much larger than
the masses of the new gauge bosons $W'$ and $Z'$. Nevertheless,
the structure of LRSM typically arises as a low-energy effective theory
for most grand unified theories---supersymmetric or not---whose
gauge symmetry breaking pattern includes the gauge
group $SO(10)$.

In this note, the sensitivity offered by the relative longitudinal
polarisation measurement mentioned above is considered\cite{Quin}
in the case of allowed $\beta$-decays. Specific results
are established for mirror nuclei in the context of {\em general\/}
LRSM, following the discussion
of Ref.\cite{Oscar} which analysed the sensitivity offered by
{\em asymmetry\/} measurements in the context of so-called
Manifest Left-Right Symmetric Models (MLRSM) which are constructed such that
at tree-level gauge coupling constants and flavour mixing matrices be identical
for both chirality sectors of the theory.
In fact, two such polarisation-asymmetry correlation experiments have already
yielded results in the cases of the allowed pure Gamow-Teller (GT)
decays of $^{107}$In\cite{107In} and $^{12}$N\cite{12N},
for which the sen\-si\-ti\-vi\-ty to possible right-handed
charged current contributions must {\em a priori\/} be among the largest
attainable, owing to the Gamow-Teller character of these transitions
and their large asymmetry parameter.

The outline of the note is as follows. Sect.\ref{Sect2} addresses the
longitudinal polarisation-asymmetry correlation measurement in the general case
of allowed $\beta$-decays assuming only vector and axial contributions
to the charged current interaction. In Sect.\ref{Sect3}, these
results are particularised to {\em general\/}
Left-Right Symmetric Models.
The case of mirror nuclei is then
analysed in Sect.\ref{Sect4}, in order to identify possible
attractive candidates for this type of measurements beyond the
cases of allowed pure GT decays. This is done
on general grounds, without paying attention to the necessary
considerations concerning the technical
feasibility either of such experiments or of the required degree
of nuclear polarisation, which would have to be addressed
on a case by case basis. Further results
of interest are presented in
an Appendix, while the note ends with some general conclusions.

\section{Allowed $\beta$-decay and $(V,A)$ charged currents}
\label{Sect2}

The expressions for physical observables of relevance
to allowed $\beta$-decay in the
case of the most general four-fermi effective interaction are
available from Ref.\cite{JTW}. In this note, {\em only vector $(V)$ and
axial $(A)$ a priori complex four-fermi
coupling coefficients are considered\/}, associated
to pure vector and axial charged current contributions to $\beta$-decay.
{\em Under this specific restriction\/} and ignoring recoil
order corrections\footnote{In the case of
super-allowed decays such as those of mirror nuclei, these corrections
are small and such an approximation is certainly justified
in the present context which aims at identifying potential
candidates for relative longitudinal polarisation-asymmetry
correlation experiments.},
when only the $\beta^{\mp}$ particle is
observed the corresponding distribution is then given by the spin density
matrix\cite{JTW}
\begin{displaymath}
\frac{d^3W}{dE\,d^2\Omega}\ =\ W_0(E)\,\xi\,\Bigg\{\ 1\,+\,\beta J A
(\hat{p}.\hat{J})\ +
\end{displaymath}
\begin{equation}
+\ \vec{\sigma}.\Bigg[\beta G\hat{p}+J\frac{\gamma_z}{\gamma}N'\hat{J}+
J(1-\frac{\gamma_z}{\gamma})N'(\hat{p}.\hat{J})\hat{p}
\mp J\frac{\alpha Z}{\gamma} A (\hat{J}\times\hat{p})\Bigg]\Bigg\}\ \ ,
\label{eq:Dist1}
\end{equation}
with the following notation. The parameter $(-1\le J\le +1)$
stands for the degree
of nuclear polarisation of the oriented nucleus, while the normalised
vector $\hat{J}$ represents the direction of nuclear polarisation.
Of course, $\beta$ is the velocity of the emitted $\beta^{\mp}$
particle---normalised to the speed of light in vacuum---, with $\gamma$
the associated relativistic dilatation factor $(\gamma=1/\sqrt{1-\beta^2})$,
while $E$, $\hat{p}$ and $\vec{\sigma}$
are the $\beta^{\mp}$ particle total energy, normalised
momentum and polarisation (spin) quantum operator, respectively. Coulomb
corrections\cite{JTW} induce a dependence on the fine structure constant
$\alpha$, while $\gamma_z$ is defined as $(\gamma_z=\sqrt{1-(\alpha Z)^2}\,)$
where $Z$ is the charge of the {\em daughter\/} nucleus. The coefficients
$W_0(E)$, $\xi$, $A$, $G$ and $N'$ are given in the Appendix for
$\beta$-transitions of initial nuclear spin $J$ and final
nuclear spin $J'$ in terms
of the effective four-fermi complex coupling coefficients $C_V$,
$C_V'$, $C_A$ and $C_A'$ introduced in Ref.\cite{JTW}.
Finally, the choice of sign in front of the last term in the above expression
corresponds to whether it is an electron or a positron which is emitted;
throughout the note, whenever such a choice is indicated, the upper
sign always corresponds to electron emission, and the lower sign to positron
emission.

Since the purpose of the present analysis is the identification of
attractive candidates for polarisation-asymmetry correlation
experiments besides allowed pure GT decays, let us consider the ideal
situation in which $\beta^{\mp}$ particles of a specific
energy $E$---or velocity $\beta$---and initial momentum direction $\hat{p}$
are observed and for which it is {\em exactly\/}
their {\em longitudinal
polarisation\/} which is measured. All other things being equal,
dif\-fe\-rent nuclei may then be compared on an equal footing under
these identical {\em idealised\/} experimental conditions.
Obviously on a practical level,
other considerations would also have to be addressed, such as the
finite energy and angular acceptance of any experimental set-up
leading to additional contributions from transverse spin components as well,
the spin precession in magnetic fields of spectrometers or polarimeters,
and more importantly, the
feasibility of the production and of the polarisation of a given
nucleus. As mentioned in the Introduction, such issues are not
tackled in this note, since they can only be considered
on a case by case
basis once a potential candidate is identified.

Under the stated idealised conditions,
and given a nucleus of polarisation $J$, the
experimental asymmetry of the $\beta$-decay distribution is of the form
\begin{equation}
N_\beta(J)\ =\ N_0\,\Big[\,1+\beta J A (\hat{p}.\hat{J})\,\Big]\ \ ,
\end{equation}
where $N_0$ represents the normalised source activity.
Similarly, from the spin density matrix in (\ref{eq:Dist1}),
the {\em longitudinal\/} polarisation of the produced $\beta^{\mp}$
particle is then given by the expression
\begin{equation}
P_L(J)\ =\ \frac{\beta G + J N' (\hat{p}.\hat{J})}
{1+\beta J A (\hat{p}.\hat{J})}\ \ \ .
\end{equation}

Following the suggestion of Ref.\cite{Quin}, let us consider
{\em relative\/} measurements of such observables for different
values of the degree $J$ of the
polarisation of the oriented nucleus, without changing the
direction $\hat{J}$ of its orientation. The obvious advantage of such
measurements is that they are much less sensitive to systematic
effects than are absolute measurements of asymmetries or
polarisations. Therefore, given two different degrees
$J_1$ and $J_2$ of nuclear polarisation
and {\em identical\/} normalised source
activities $N_0$, the relative {\em experimental\/} asymmetry
$A_{\rm exp}(J_2,J_1)$ may be defined according to
\begin{equation}
1\ -\ A_{\rm exp}(J_2,J_1)\ =\ \frac{N_\beta(J_2)}{N_\beta(J_1)}\ =\
\frac{1+\beta J_2 A (\hat{p}.\hat{J})}
{1+\beta J_1 A (\hat{p}.\hat{J})}\ \ \ ,
\end{equation}
while the relative longitudinal polarisation is simply
\begin{equation}
R(J_2,J_1)\ =\ \frac{P_L(J_2)}{P_L(J_1)}\ =\
\frac{1}{1-A_{\rm exp}(J_2,J_1)}\ \frac{\beta G + J_2 N' (\hat{p}.\hat{J})}
{\beta G + J_1 N' (\hat{p}.\hat{J})}\ \ \ .
\end{equation}

Since the purpose of the present discussion is the identification
of potential attractive candidates for this type of measurement,
all other things being equal, it proves simpler to consider henceforth
only two specific situations with regards to nuclear polarisation.
The first is obtained when the reference nuclear polarisation $J_1$
is vanishing, $(J_1=0)$, and when $(J_2=-J)$. The second\cite{Quin}
corresponds to a situation in which
the $\beta^{\mp}$ longitudinal polarisation is considered for opposite
directions of nuclear polarisation, namely for $(J_2=-J_1=-J)$.

In the first instance, the experimental asymmetry is given by
\begin{equation}
A_{\rm exp}(-J,0)\ =\ \beta\,J\,A\,(\hat{p}.\hat{J})\ \ \ ,
\label{eq:Aexp1}
\end{equation}
which upon substitution in the expression for the relative
longitudinal polarisation $R(-J,0)$ leads to the result,
\begin{equation}
R(-J,0)\ =\ \frac{1}{\beta^2}\,\frac{1}{1-A_{\rm exp}(-J,0)}\,
\Bigg[\,\beta^2\ -\ A_{\rm exp}(-J,0)\,
\frac{\xi\,(\xi N')}{(\xi A)\,(\xi G)}\,\Bigg]\ \ \ .
\label{eq:R(-J,0)}
\end{equation}

Similarly in the second instance when $\left(\,(J_2,J_1)=(-J,J)\,\right)$,
one has
\begin{equation}
A_{\rm exp}(-J,J)\ =\ \frac{2\,\beta\,J\,A\,(\hat{p}.\hat{J})}
{1\,+\,\beta\,J\,A\,(\hat{p}.\hat{J})}\ \ \ ,
\label{eq:Aexp2}
\end{equation}
or equivalently,
\begin{equation}
\beta\,J\,A\,(\hat{p}.\hat{J})\ =\ \frac{A_{\rm exp}(-J,J)}
{2\,-\,A_{\rm exp}(-J,J)}\ \ \ .
\end{equation}
When substituted in the definition for $R(-J,J)$, one then
derives the expression for the relative longitudinal polarisation
in this case,
\begin{equation}
R(-J,J)\ =\ \frac{1}{1-A_{\rm exp}(-J,J)}\,
\Bigg[\,1\ -\ 2\ \frac{1}{1\,+\,\beta^2\,\frac{2-A_{\rm exp}(-J,J)}
{A_{\rm exp}(-J,J)}\,\frac{(\xi A)(\xi G)}{\xi (\xi N')}}\,\Bigg]\ \ \ .
\label{eq:R(-J,J)}
\end{equation}

Note that it is the same
quantity
\begin{equation}
\frac{\xi\,(\xi N')}{(\xi A)\,(\xi G)}\ \ \ ,
\label{eq:ratio}
\end{equation}
depending on the underlying physics, which appears in
the expressions for $R(-J,0)$ and $R(-J,J)$. Since
this combination of parameters takes the value unity in the SM
(see the Appendix), any deviation
of the ratio (\ref{eq:ratio}) from unity must stem from some new
physics\footnote{Note that this remark does not account for possible
recoil order corrections to the relative longitudinal
polarisation, which are ignored in this note.}
beyond the SM. Let us thus introduce the quantity
$\Delta$ defined by\footnote{The numerical factor of a quarter is for later
convenience.}
\begin{equation}
\Delta\ \equiv\ \frac{1}{4}\,\Big[\,
\frac{\xi\,(\xi N')}{(\xi A)\,(\xi G)}\ -\ 1\,\Big]\ =\
\frac{1}{4}\,\frac{\xi(\xi N')-(\xi A)(\xi G)}{(\xi A)(\xi G)}\ \ \ ,
\end{equation}
which thus characterises the new physics beyond the SM which
may be probed through
relative longitudinal polarisation-asymmetry correlation experiments.

In order to assess the sensitivity of these measurements
to such new physics, it
is necessary to compare the values taken by the relative
polarisations $R(-J,0)$ and $R(-J,J)$ to their values
obtained simply by setting $(\Delta=0)$ in (\ref{eq:R(-J,0)})
and (\ref{eq:R(-J,J)}) without modification of the
{\em experimental\/} asymmetries $A_{\rm exp}$.
This leads to, respectively,
\begin{equation}
R_0(-J,0)\ =\ \frac{1}{\beta^2}\,\frac{\beta^2\,-\,A_{\rm exp}(-J,0)}
{1\,-\,A_{\rm exp}(-J,0)}\ \ \ ,
\end{equation}
and
\begin{equation}
R_0(-J,J)\ =\ \frac{1}{1\,-\,A_{\rm exp}(-J,J)}\
\frac{\beta^2\Big[2-A_{\rm exp}(-J,J)\Big]\,-\,
A_{\rm exp}(-J,J)}
{\beta^2\Big[2-A_{\rm exp}(-J,J)\Big]\,+\,A_{\rm exp}(-J,J)}\ \ \ .
\end{equation}
Let us emphasize that these expressions {\em are not
those which one may derive in the SM\/} for the
relative longitudinal polarisations,
since they are obtained simply by setting only the quantity
$\Delta$ to zero in the results for $R(-J,0)$ and $R(-J,J)$
without assuming that the asymmetry parameter $A$ is given
by its value $A_0$ in the SM.
Indeed, the expressions for $R_0(-J,0)$
and $R_0(-J,J)$ still involve the {\em experimental\/} asymmetries
$A_{\rm exp}(-J,0)$ and $A_{\rm exp}(-J,J)$
which may differ from the values predicted by the SM
if new physics does contribute to the asymmetry parameter $A$,
as made explicit in (\ref{eq:Aexp1}) and (\ref{eq:Aexp2}).
Nevertheless, it is obviously useful\cite{Quin} to express
the relative longitudinal polarisation in terms of the {\em directly
observable\/} experimental asymmetries $A_{\rm exp}(-J,0)$
and $A_{\rm exp}(-J,J)$.

Given the quantities $R_0(-J,0)$ and $R_0(-J,J)$, any genuine physical
deviation from the SM contributing to $R(-J,0)$ and $R(-J,J)$ would
be manifested through a value different from unity for the ratios,
\begin{equation}
\frac{R(-J,0)}{R_0(-J,0)}\ \ \ ,\ \ \
\frac{R(-J,J)}{R_0(-J,J)}\ \ \ .
\end{equation}
In terms of the quantities introduced in the forthcoming definitions,
it is straightforward to establish the following {\em exact\/}
results, valid independently of whether the parameter $\Delta$
is small in comparison to unity or not. One finds,
\begin{equation}
\frac{R(-J,0)}{R_0(-J,0)}\ =\ 1\ -\ k(-J,0)\,\Delta\ \ \ ,
\end{equation}
with the factor $k(-J,0)$ given by
\begin{equation}
k(-J,0)\ =\ 4\,\frac{A_{\rm exp}(-J,0)}
{\beta^2\,-\,A_{\rm exp}(-J,0)}\ \ \ .
\end{equation}
Similarly for $R(-J,J)$, one has
\begin{equation}
\frac{R(-J,J)}{R_0(-J,J)}\ =\ 1\ -\ k(-J,J)\,
\frac{\Delta}{1+4\,\frac{A_{\rm exp}(-J,J)}{\beta^2
\left[2-A_{\rm exp}(-J,J)\right]
+ A_{\rm exp}(-J,J)}\,\Delta}\ \ \ ,
\label{eq:RR02}
\end{equation}
with the factor $k(-J,J)$ defined by
\begin{equation}
k(-J,J)\ =\ 8\ \frac{\beta^2\,A_{\rm exp}(-J,J)\,
\Big[2-A_{\rm exp}(-J,J)\Big]}
{\beta^4\,\Big[2-A_{\rm exp}(-J,J)\Big]^2-A^2_{\rm exp}(-J,J)}\ \ \ .
\end{equation}

Note that under the considered idealised situation in which it is
exactly the longitudinal $\beta^{\mp}$ polarisation which is measured,
both the values of $R_0(-J,0)$ or $R_0(-J,J)$ and of the
factors $k(-J,0)$ or $k(-J,J)$
only depend\cite{Quin}
on the corresponding
{\em observed experimental\/} asymmetry $A_{\rm exp}$
and on the value of $\beta^2$.
Moreover, the factors $k(-J,0)$ and $k(-J,J)$
offer an enhanced\cite{Quin}
sensitivity to any deviation from the value unity expected in the SM
for the corresponding ratio $R/R_0$ of relative longitudinal polarisations.
Indeed, the factor appropriate to the first case, namely $k(-J,0)$,
diverges\footnote{This divergence does not entail a loss of
physical significance of the results, but follows simply from the fact
that the quantity $R_0(-J,0)$ vanishes as the experimental
asymmetry $A_{\rm exp}(-J,0)$ approaches the value $A_{\rm exp}^{(0)}(-J,0)$,
while at the same time the product $R_0(-J,0)\,k(-J,0)$ remains
finite as it should since $R(-J,0)$ is finite under all circumstances.
In other words, the significance of the divergence is that
when the experimental asymmetry $A_{\rm exp}(-J,0)$ is
optimised at the value $A^{(0)}_{\rm exp}(-J,0)$, the
contribution of $(-R_0(-J,0)\,k(-J,0)\,\Delta)$
to $R(-J,0)$ becomes increasingly larger than that of $R_0(-J,0)$.
Of course, the same comments also apply to the case of $R(-J,J)$
which follows.}
as the experimental asymmetry $A_{\rm exp}(-J,0)$ approches
the value
\begin{equation}
A_{\rm exp}^{(0)}(-J,0)\ =\ \beta^2\ \ .
\end{equation}
Similarly, the enhancement factor appropriate
to the second case, namely $k(-J,J)$,
diverges as the experimental asymmetry $A_{\rm exp}(-J,J)$
approches the value
\begin{equation}
A_{\rm exp}^{(0)}(-J,J)\ =\ \frac{2\,\beta^2}{1+\beta^2}\ \ ,
\end{equation}
while the quantity multiplied by $k(-J,J)$ in (\ref{eq:RR02}) then
reduces to
\begin{equation}
\frac{\Delta}{1+2\Delta}\ \ \ .
\end{equation}
Note that in either case, the optimal experimental asymmetry
$A^{(0)}_{\rm exp}$ corresponds to a degree of nuclear
polarisation $J$ and a choice of $\beta$ such that
\begin{equation}
A\,J\,(\hat{p}.\hat{J})\ =\ \beta\ \ \ .
\label{eq:bJA}
\end{equation}
In other words, for a given nucleus, namely a given asymmetry
parameter $A$, the optimal sensitivity
to a possible contribution
from right-handed currents is achieved for values of the
{\em effective\/} degree of nuclear polarisation,
\begin{equation}
{\cal P}\ =\ |J(\hat{p}.\hat{J})|\ \ \ ,
\end{equation}
and of the $\beta^\mp$ particle velocity $\beta$ such that
\begin{equation}
|A\,{\cal P}|\ =\ \beta\ \ \ ,
\label{eq:optimal}
\end{equation}
the choice of sign for $J(\hat{p}.\hat{J})$
being such that the experimental asymmetries
$A_{\rm exp}$ as defined in this note be positive.

These conclusions correspond to the advertised\cite{Quin} sensitivity of this
type of measurement to right-handed currents: the closer the experimental
asymmetry $A_{\rm exp}$ to $A_{\rm exp}^{(0)}$, namely the closer
the choice of values of $(\beta,{\cal P})$ to the optimal situation
such that $(\beta=|A{\cal P}|)$, the
larger the sensitivity to a possible
deviation $(\Delta\ne 0)$ from the SM. Since values of $\beta$
for which the decay count rate is the largest
are typically close to the maximal value of unity\footnote{Small,
albeit vanishing values of $\beta$ are of course possible also,
in which case the optimal sensitivity is achieved for small
effective nuclear polarisations ${\cal P}$. However, $\beta^{\mp}$ count
rates decrease as $\beta$ approaches zero, thus leading to a loss
in statistics for any precision measurement.}, clearly the best
sensitivity requires both the largest possible effective degree of
nuclear polarisation ${\cal P}$ and the largest possible
asymmetry parameter $|A|$. In particular, since the asymmetry
parameter $|A|$ for allowed pure GT transitions of nuclear spin sequence
$(J'=J-1)$ is maximal in the SM
(see the Appendix), such $\beta$-decays
are certainly among the best candidates to probe for right-handed
currents through longitudinal polarisation-asymmetry correlation experiments.
This is the case for example
for the $^{107}$In\cite{107In} and $^{12}$N\cite{12N} nuclei.

Given the expressions for $\xi$, $A$, $G$ and $N'$
listed in the Appendix, the quantity $\Delta$ may easily be
related to the underlying effective coupling coefficients
$C_V$, $C_V'$, $C_A$ and $C_A'$. It proves useful to introduce
the notation,
\begin{equation}
a_L\ =\ M_F^2\,|C_V+C_V'|^2\ +\ M^2_{GT}\,|C_A+C_A'|^2\ \ \ ,
\label{eq:aL}
\end{equation}
\vspace{3pt}
\begin{equation}
a_R\ =\ M_F^2\,|C_V-C_V'|^2\ +\ M^2_{GT}\,|C_A-C_A'|^2\ \ \ ,
\label{eq:aR}
\end{equation}
\vspace{3pt}
\begin{equation}
b_L\ =\ \mp M^2_{GT}\,\lambda_{J'J}\,|C_A+C_A'|^2\ -\
2\delta_{J'J} M_F M_{GT}\,\sqrt{\frac{J}{J+1}}\,{\rm Re}
\left((C_V+C_V')(C_A^*+{C_A'}^*)\right)\ \ \ ,
\label{eq:bL}
\end{equation}
\begin{equation}
b_R\ =\ \mp M^2_{GT}\,\lambda_{J'J}\,|C_A-C_A'|^2\ -\
2\delta_{J'J} M_F M_{GT}\,\sqrt{\frac{J}{J+1}}\,{\rm Re}
\left((C_V-C_V')(C_A^*-{C_A'}^*)\right)\ \ \ ,
\label{eq:bR}
\end{equation}
in terms of which one then derives the {\em exact\/}
expression
\begin{equation}
\Delta\ =\ \frac{1}{2}\,\frac{a_L\,b_R\ +\ a_R\,b_L}
{(a_L\,-\,a_R)(b_L\,-\,b_R)}\ \ \ .
\label{eq:4Delta}
\end{equation}
Since the numerator of the result in (\ref{eq:4Delta}) involves
precisely the differences $(C_V-C_V')$ and $(C_A-C_A')$
which vanish for purely left-handed couplings as is the
case in the SM, it is
clear that the quantity $\Delta$---probed through
relative longitudinal polarisation-asymmetry correlation measurements---is
indeed sensitive to
right-handed charged current contributions to allowed $\beta$-decay.

Note also that in terms of the quantities $a_L$, $a_R$, $b_L$ and $b_R$
introduced above, the asymmetry parameter $A$ reads,
\begin{equation}
A\ =\ \frac{b_L\,-\,b_R}{a_L\,+\,a_R}\ \ \ .
\end{equation}

When deviations of the coefficients $C_V$, $C_V'$, $C_A$
and $C_A'$ from their values in the SM are small, it is justified to
consider a first order expansion of $\Delta$ in the quantities
$a_R$ and $b_R$, leading to
\begin{equation}
\Delta\ \simeq\ \frac{1}{2}\,\Bigg[\,\frac{a_R}{(a_L)_0}\ +\
\frac{b_R}{(b_L)_0}\,\Bigg]\ \ \ ,
\end{equation}
where $(a_L)_0$ and $(b_L)_0$ are the values of $a_L$ and $b_L$
in the SM, respectively,
\begin{equation}
(a_L)_0\ =\ 4\,|C_V^{(0)}|^2\,M_F^2\,\Big[\,1+\lambda^2\,\Big]\ \ \ ,
\end{equation}
and
\begin{equation}
(b_L)_0\ =\ 4\,|C_V^{(0)}|^2\,M_F^2\,\Big[\,
\mp\,\lambda^2\,\lambda_{J'J}\ -\ 2\,\delta_{J'J}\lambda\,
\sqrt{\frac{J}{J+1}}\,\Big]\ \ \ .
\end{equation}
Here, $C_V^{(0)}$ is the value of the coefficient $C_V$ in the
SM, while $\lambda$ is the ratio
\begin{equation}
\lambda\ =\ \frac{g_A}{g_V}\,\frac{M_{GT}}{M_F}\ \ \ ,
\label{eq:lambda}
\end{equation}
$g_V$ and $g_A$ being the nucleon vector and axial couplings,
respectively (see the
Appendix for further details).

Another situation of particular interest is that of
allowed pure GT transitions, for which one simply finds,
\begin{equation}
A_{|_{GT}}\ =\ {A_0}_{|_{GT}}\,
\frac{|C_A+C_A'|^2\,-\,|C_A-C_A'|^2}
{|C_A+C_A'|^2\,+\,|C_A-C_A'|^2}\ \ \ ,
\label{eq:AGT}
\end{equation}
${A_0}_{|_{GT}}$ being the asymmetry parameter for allowed pure GT
decays in the SM (see the Appendix), as well as
\begin{equation}
\Delta_{|_{GT}}\ =\ \frac{1}{4}\,\Bigg[
\left(\frac{{A_0}_{|_{GT}}}{A_{|_{GT}}}\right)^2\,-\,1\Bigg]\ =\
\frac{|C_A-C_A'|^2}{|C_A+C_A'|^2}\,
\frac{1}{\Big[1-\frac{|C_A-C_A'|^2}{|C_A+C_A'|^2}\Big]^2}\ \ \ .
\end{equation}
These expressions are independent of the nucleus involved
and of the specifics of the
underlying new physics which would be leading to $(\Delta_{|_{GT}}\ne 0)$.
Thus in such a case, the parameter $\Delta$ indeed provides a {\em direct\/}
measure for right-handed current contributions to the coupling of
the leptonic charged current to the hadronic axial
charged current in nuclear $\beta$-decay. On the other hand, note that
even though the quantity $\Delta$ is in fact related
to the asymmetry parameter $A$ in the case of allowed pure GT transitions,
a measurement of the relative longitudinal polarisation-asymmetry
correlation is potentially far more sensitive
to contributions from right-handed
currents than is a measurement of the asymmetry parameter $A$
itself. Indeed,
the former sensitivity to the ratio
$\Big(|C_A-C_A'|^2/|C_A+C_A'|^2\Big)$
is characterised by the enhancement factors
$k(-J,0)$ and $k(-J,J)$ which {\em a priori\/} may reach quite large
values by appropriate choices of $\beta$ and of the effective degree
of nuclear polarisation ${\cal P}$. On the other hand,
the sensitivity
of the asymmetry parameter $A$ to the same ratio
is essentially characterised by a
fixed enhancement factor of two only, as follows from (\ref{eq:AGT}).

Before concluding this general discussion, let us address one last
issue. As was already pointed out previously,
since in most cases the value of $\beta$ is not much different
from unity in the energy domain where the $\beta^\mp$ count
rate is maximal, for a given effective degree of nuclear polarisation
${\cal P}$ the enhancement factors $k(-J,0)$ and $k(-J,J)$
are the largest for experimental asymmetries $A_{\rm exp}(-J,0)$
and $A_{\rm exp}(-J,J)$ as
close to the value unity as possible. However, this also implies
that the count rate of $\beta^{\mp}$ particles associated to the
direction of nuclear polarisation for which the sensitivity to $\Delta$ is
the largest, is also the smaller the closer the experimental
asymmetry to the value unity. Indeed, this count rate
at the optimal sensitivity such that $(|A{\cal P}|=\beta)$
is proportional to $(1-\beta^2)$. Nevertheless, it is possible to show
that the loss in count rate is compensated for by the
gain in sensitivity; for either configuration of nuclear
polarisations considered in this note, the figure of merit characterising the
precision with which a deviation from the value unity for the
ratio $R/R_0$ may be established experimentally
is indeed optimal for
the previously given value $A_{\rm exp}^{(0)}$ of the
experimental asymmetry $A_{\rm exp}$ at which the corresponding
enhancement factor $k$ is the largest.

\section{General Left-Right Symmetric Models}
\label{Sect3}

The results of the previous section are valid quite generally for allowed
decays,
since the only assumptions made so far are that the
effective Hamiltonian for $\beta$-decay receives contributions
from vector and axial
currents only, with arbitrary complex coupling coefficients,
and that recoil order corrections to relative
longitudinal polarisations are negligible.
Let us now particularise the discussion to {\em general\/} Left-Right
Symmetric Models, based on the gauge group
$SU(2)_L\times SU(2)_R\times U(1)_{B-L}$ in the electroweak sector.
In as far as semi-leptonic charged weak interactions are
concerned, contributions to the $\beta$-decay process in such models
follow from charged gauge boson and Higgs exchanges. However,
charged Higgs exchanges---when contributing\footnote{Charged Higgs
exchanges do not contribute in the case of allowed pure GT
transitions.}---shall be ignored in
the present discussion, assuming that they are suppressed
through small coupling constants and large masses.
This effectively
leaves only charged gauge boson exchanges,
namely those of the ordinary gauge boson $W$ of mass\cite{PDG}
\begin{equation}
M_1\ =\ 80.22\,\pm\,0.26\ {\rm GeV}/c^2\ \ ,
\end{equation}
and of the hypothetical heavy charged gauge boson $W'$ of mass $M_2$.
Thus, since scalar and pseudoscalar charged Higgs exchange contributions are
ignored, within the framework of LRSM $\beta$-decay processes
indeed receive contributions from vector and axial couplings only,
namely from left- and right-handed fermionic gauge currents.

However, the propagating gauge bosons $W$ and $W'$ are not necessarily those
which couple to fermions of definite chirality. Indeed, the physical
charged gauge bosons $W$ and $W'$ and the charged gauge bosons associated
to the underlying gauge group $SU(2)_L\times SU(2)_R$ which thus
couple to currents of specific chirality, are related to
one another through the mixing matrix\cite{Herczeg},
\begin{equation}
\begin{array}{c c l}
W_L^+ & = & \cos\zeta\,W_1^+\ +\ \sin\zeta\,W_2^+\ \ \ ,\\ \\
W_R^+ & = & e^{i\omega}\Big[\,-\sin\zeta\,W_1^+\ +\
\cos\zeta\,W_2^+\,\Big]\ \ \ ,
\end{array}
\end{equation}
or equivalently
\begin{equation}
\begin{array}{c c l}
W_1^+ & = & \cos\zeta\,W_L^+\ -\ e^{-i\omega}\sin\zeta\,W_R^+\ \ \ ,\\ \\
W_2^+ & = & \sin\zeta\,W_L^+\ +\ e^{-i\omega}\cos\zeta\,W_R^+\ \ \ .
\end{array}
\end{equation}
Here, $W_L$ and $W_R$ denote the fundamental gauge bosons
coupling to the fermionic currents of left- and right-handed chirality,
respectively, while $W_1$ and $W_2$ denote the physical mass eigenstate
gauge bosons of masses $M_1$ and $M_2$, respectively.
The parameter $\zeta$ is a mixing angle\footnote{Our choice
of sign for $\zeta$ is opposite to that made in Ref.\cite{Lang}
but agrees with that of Ref.\cite{Herczeg}.} for charged gauge
bosons, while the parameter $\omega$ determines a CP violating
phase originating from complex vacuum expectation values in the
Higgs sector. These quantities are constrained
phenomenologically in certain classes of LRSM\cite{Herczeg,Lang}.

In addition, the coupling strength of the gauge bosons $W_L$ and $W_R$ to the
fundamental fermions is specified by the gauge coupling constants
$g_L$ and $g_R$, respectively. Again,
the ratio $g_R/g_L$ is constrained
phenomenologically\cite{Cvetic}.

Finally, the coupling of the charged gauge bosons $W_L$ and $W_R$
to quarks and leptons also involves different Cabibbo-Kobayashi-Maskawa (CKM)
flavour mixing matrices in each chirality sector. Since in the
hadronic sector, only the up and down quarks couple to the
$\beta$-decay process, the relevant CKM matrix elements are
denoted
\begin{equation}
V^L_{ud}\ \ \ ,\ \ \ V^R_{ud}\ \ \ ,
\end{equation}
for the left- and right-handed sectors, respectively.
Similarly in the leptonic sector, {\em a priori\/} the emitted
electron or positron may be produced together
with a mass eigenstate neutrino $\nu_i$,
with an amplitude determined by leptonic
CKM matrix elements denoted as
\begin{equation}
U^L_{ie}\ \ \ ,\ \ \ U^R_{ie}\ \ \ .
\end{equation}
Generally, the ratios
\begin{equation}
v_{ud}\ =\ \frac{V^R_{ud}}{V^L_{ud}}\ \ \ ,\ \ \
v_{ie}\ =\ \frac{U^R_{ie}}{U^L_{ie}}\ \ \ ,\ \ \
\label{eq:comb1}
\end{equation}
are arbitrary complex numbers, related to the underlying complex Yukawa
couplings and Higgs vacuum expectation values, thus
potentially leading to new
CP violating processes in their own right. Indeed,
even though it is always possible by an appropriate choice of
phases of the fermionic fields to fix the quark as well as the
leptonic CKM matrix elements $V^{(L,R)}_{ud}$ and $U^{(L,R)}_{ie}$
either in the left- or in the right-handed sectors
to be real---as is the case for the ordinary Cabibbo angle---, this is
not possible for both sectors simultaneously. Assuming that either
$v_{ud}$ or $v_{ie}$ or both be real, would imply particular restrictions
on the class of LRSM being considered. Here again,
there exist\cite{Herczeg,Lang}
certain phenomenological constraints on these quark and lepton CKM matrix
elements.

The above description thus provides the complete set of parameters
required for the application of the discussion of Sect.\ref{Sect2}
to {\em general\/} LRSM, no assumption nor approximation
whatsoever as to the definition of such models
being implied at this stage.
It proves useful to introduce the following combinations of these
quantities,
\begin{equation}
t\ =\ \tan\zeta\ \ \ ,\ \ \
\delta\ =\ \frac{M^2_1}{M^2_2}\ \ \ ,\ \ \
r\ =\ \frac{g_R}{g_L}\ \ ,
\label{eq:comb2}
\end{equation}
\begin{equation}
v_u\ =\ \frac{|V^R_{ud}|^2}{|V^L_{ud}|^2}\ =\ |v_{ud}|^2\ \ \ ,\ \ \
v_e\ =\ \frac{\sum_i'|U^R_{ie}|^2}{\sum_i'|U^L_{ie}|^2}\ \ \ ,
\label{eq:comb3}
\end{equation}
and finally
\begin{equation}
\eta_0\ =\ \frac{1}{M^2_1}\left(\frac{g_L^2}{8}\right)\,\cos^2\zeta\ =\
\frac{1}{M^2_1}\left(\frac{g_L^2}{8}\right)\,\frac{1}{1+t^2}\ \ \ .
\label{eq:comb4}
\end{equation}
In particular, the symbol $\sum_i'$
appearing in the definition
of the quantity $v_e$ in a notation to be detailed
presently stems from the following fact.
Since the neutrino produced in the $\beta$-decay process is not
observed, any measurement involves a sum over all neutrino mass
eigenstates whose production is not forbidden kinematically.
Therefore, assuming that all neutrinos produced in the process
have a mass sufficiently small in order not to induce
a significant distortion of the $\beta^{\mp}$ energy distribution,
one need only sum the
corresponding decay spectra over all such mass eigenstate neutrinos
$\nu_i$ without accounting for a modification
in phase-space factors. In other words, the symbol $\sum_i'$ stands
for a sum over all neutrinos $\nu_i$ whose production is
not kinematically suppressed. In particular, note that when all
mass eigenstate neutrinos
do participate in the process, each of the sums
\begin{equation}
{\sum_i}'|U^L_{ie}|^2\ \ \ ,\ \ \
{\sum_i}'|U^R_{ie}|^2\ \ \ ,
\end{equation}
then reduces to the value unity, owing to the unitarity of the
corresponding leptonic CKM flavour mixing matrix\cite{Herczeg},
in which case one simply has $(v_e=1)$.

So-called Manifest Left-Right Symmetric Models (MLRSM) are those
LRSM such that the gauge coupling constants $g_R$ and $g_L$
and the quark and lepton CKM matrices in the left- and right-handed
sectors are equal, and such that
the CP violating phase
$\omega$ vanishes. Namely, these MLRSM are such that except for the different
masses for the charged $W$ and $W'$ and neutral $Z$ and $Z'$ gauge bosons,
the sectors of left- and right-handed chirality are indistinguishable,
and no CP violation originates in these models except for the
two ordinary Kobayashi-Maskawa phases appearing in
quark and lepton CKM matrices
associated to three generations of quarks and leptons.
In particular, parity invariance is then effectively restored
in these MLRSM for
processes of high momenta transfers.
The present purely experimental lower limit on the mass $M_2$
of a charged heavy gauge boson $W'$ in the context of these MLRSM
is\cite{CDF},
\begin{equation}
M_2\ >\ 652\ {\rm GeV}/c^2\ \ \ (95\%\ {\rm C.L.})\ \ .
\end{equation}
However, let us emphasize here again that the restrictions leading
to MLRSM are not assumed
in this note; our discussion applies to the most general LRSM possible.

Given this description of weak charged current interactions in
general LRSM,
it is straightforward to determine the corresponding
four-fermi effective Hamiltonian\cite{Herczeg,Lang}
at the quark-lepton level relevant to $\beta$-decay, which
is thus of the form,

\pagebreak

\begin{displaymath}
H^{\rm quark}_{\rm eff}\ ={\hspace{380pt}}
\end{displaymath}
\begin{equation}
\begin{array}{r l}
=&\eta_{LL}\overline{\psi}_u\gamma_\mu(1-\gamma_5)\psi_d\,
	 \overline{\psi}_e\gamma^\mu(1-\gamma_5)\psi_{\nu_e}\ +\
\eta_{LR}\overline{\psi}_u\gamma_\mu(1-\gamma_5)\psi_d\,
	 \overline{\psi}_e\gamma^\mu(1+\gamma_5)\psi_{\nu_e}\ + \\ \\
+&\eta_{RL}\overline{\psi}_u\gamma_\mu(1+\gamma_5)\psi_d\,
	 \overline{\psi}_e\gamma^\mu(1-\gamma_5)\psi_{\nu_e}+
\eta_{RR}\overline{\psi}_u\gamma_\mu(1+\gamma_5)\psi_d\,
	 \overline{\psi}_e\gamma^\mu(1+\gamma_5)\psi_{\nu_e}\ \ ,
\end{array}
\end{equation}
$\psi$ denoting the ordinary Dirac spinors for quarks and leptons.
The coefficients $\eta_{LL}$, $\eta_{LR}$, $\eta_{RL}$ and
$\eta_{RR}$ are given by
\begin{equation}
\begin{array}{r c l}
\eta_{LL}&=&\ \,\eta_0\,v_{LL}\,\Big(1+\delta t^2\Big)\ \ \ ,\\ \\
\eta_{LR}&=&-\,\eta_0\,v_{LR}\,r t\,e^{-i\omega}\Big(1-\delta\Big)\ \ \ ,\\ \\
\eta_{RL}&=&-\,\eta_0\,v_{RL}\,r t\,e^{i\omega}\Big(1-\delta\Big)\ \ \ ,\\ \\
\eta_{RR}&=&\ \,\eta_0\,v_{RR}\,r^2\Big(t^2+\delta\Big)\ \ \ ,
\end{array}
\end{equation}
with the notation
\begin{equation}
v_{LL}=V^L_{ud}\,{U^L_{ie}}^*\ \ ,\ \
v_{LR}=V^L_{ud}\,{U^R_{ie}}^*\ \ ,\ \
v_{RL}=V^R_{ud}\,{U^L_{ie}}^*\ \ ,\ \
v_{RR}=V^R_{ud}\,{U^R_{ie}}^*\ \ \ .
\end{equation}
In the particular case of MLRSM, these expressions agree of course
with those derived in Refs.\cite{Beg,Hol}.

At the nucleon level, the effective four-fermi Hamiltonian is
defined in terms of the couplings $C_V$, $C_V'$, $C_A$ and $C_A'$
introduced in the Appendix. Given the relations above, in LRSM these
coupling coefficients are thus determined to be
\begin{equation}
\begin{array}{r c l}
C_V &=&g_V\,\Big[\,\ \eta_{LL}\,+\,\eta_{LR}\,
	          +\,\eta_{RL}\,+\,\eta_{RR}\,\Big]\ \ ,\\ \\
C_V'&=&g_V\,\Big[\,\ \eta_{LL}\,-\,\eta_{LR}\,
	          +\,\eta_{RL}\,-\,\eta_{RR}\,\Big]\ \ ,\\ \\
C_A &=&g_A\,\Big[\,\ \eta_{LL}\,-\,\eta_{LR}\,
	          -\,\eta_{RL}\,+\,\eta_{RR}\,\Big]\ \ ,\\ \\
C_A'&=&g_A\,\Big[\,\ \eta_{LL}\,+\,\eta_{LR}\,
	          -\,\eta_{RL}\,-\,\eta_{RR}\,\Big]\ \ .\\ \\
\end{array}
\label{eq:coeffLRSM}
\end{equation}

With the help of the results listed in the Appendix, it is
then possible to compute the expression of
any observable of interest. In particular,
the asymmetry parameter $A$ and the quantity
$\Delta$ as defined in (\ref{eq:4Delta}) are given by, respectively,
\begin{equation}
A\ =\ \frac{\mp\lambda^2\lambda_{J'J}\,\Big[\,Z_+-X_+\,\Big]\ -\
2\delta_{J'J}\lambda\sqrt{\frac{J}{J+1}}\,\Big[\,T+Y\,\Big]}
{\Big[\,Z_-+X_-\,\Big]\ +\ \lambda^2\,\Big[\,Z_++X_+\,\Big]}\ \ \ ,
\label{eq:ALRSM}
\end{equation}
and
\begin{displaymath}
\Delta\ =\ \frac{1}{2}\,\frac{1}
{\Big[(Z_- - X_-)+\lambda^2(Z_+ - X_+)\Big]\,
\Big[\mp\lambda^2\lambda_{J'J}(Z_+ - X_+)-2\delta_{J'J}\lambda
\sqrt{\frac{J}{J+1}}\,(T+Y)\Big]}\times
\end{displaymath}
\begin{equation}
\begin{array}{c l}
\times & \Bigg\{\,\mp\lambda^2\lambda_{J'J}
\Big[(X_+ Z_- +X_- Z_+)+2\lambda^2 X_+ Z_+\Big]\ -\\ \\
  &\ -\ 2\delta_{J'J}\lambda\sqrt{\frac{J}{J+1}}\,
\Big[(X_- T - Z_- Y)+\lambda^2(X_+ T - Z_+ Y)\Big]\Bigg\}\ \ \ .
\end{array}
\label{eq:DeltaLRSM}
\end{equation}
In these expressions, the parameter $\lambda$ is defined
in (\ref{eq:lambda}),
while the quantities $X_{\pm}$, $Y$, $Z_{\pm}$ and $T$ are
given by,
\begin{equation}
X_+\ =\ v_e r^2 t^2 (1-\delta)^2 + v_u v_e r^4 (t^2+\delta)^2 +
2\,{\rm Re}\left(v_{ud}e^{i\omega}\right) v_e r^3 t (1-\delta)
(t^2+\delta)\ \ \ ,
\label{eq:Xplus}
\end{equation}
\begin{equation}
X_-\ =\ v_e r^2 t^2 (1-\delta)^2 + v_u v_e r^4 (t^2+\delta)^2 -
2\,{\rm Re}\left(v_{ud}e^{i\omega}\right) v_e r^3 t (1-\delta)
(t^2+\delta)\ \ \ ,
\label{eq:Xminus}
\end{equation}
\begin{equation}
Y\ =\ v_e r^2 t^2 (1-\delta)^2 - v_u v_e r^4 (t^2+\delta)^2\ \ \ ,
\label{eq:Y}
\end{equation}
\begin{equation}
Z_+\ =\ (1+t^2\delta)^2 + v_u r^2 t^2 (1-\delta)^2 +
2\,{\rm Re}\left(v_{ud}e^{i\omega}\right) rt (1-\delta) (1+t^2\delta)\ \ ,
\label{eq:Zplus}
\end{equation}
\begin{equation}
Z_-\ =\ (1+t^2\delta)^2 + v_u r^2 t^2 (1-\delta)^2 -
2\,{\rm Re}\left(v_{ud}e^{i\omega}\right) rt (1-\delta) (1+t^2\delta)\ \ ,
\label{eq:Zminus}
\end{equation}
\begin{equation}
T\ =\ (1+t^2\delta)^2 - v_u r^2 t^2 (1-\delta)^2\ \ \ .
\label{eq:T}
\end{equation}

In particular for allowed pure GT transitions, one simply has
\begin{equation}
A_{|_{GT}}\ =\ {A_0}_{|_{GT}}\,\frac{Z_+-X_+}{Z_++X_+}\ \ \ ,\ \ \
\Delta_{|_{GT}}\ =\ \frac{1}{4}\,\Bigg[
\left(\frac{{A_0}_{|_{GT}}}{A_{|_{GT}}}\right)^2\,-\,1\Bigg]\ =\
\frac{X_+ Z_+}{\Big[Z_+ - X_+\Big]^2}\ \ \ ,
\end{equation}
${A_0}_{|_{GT}}$ being the asymmetry parameter in the SM for allowed pure GT
decays, given in (\ref{eq:Appendix.A0GT}) in the Appendix.

Another particular case of interest is obtained when the mixing
angle $\zeta$ vanishes identically, $(\zeta=0)$, for which one finds,
\begin{equation}
A_{|_{\zeta=0}}\ =\ A_0\ \frac{1-v_u v_e \, r^4\, \delta^2}
{1+v_u v_e \, r^4\, \delta^2}\ \ \ ,\ \ \
\Delta_{|_{\zeta=0}}\ =\ \frac{v_u v_e\,r^4\,\delta^2}
{\Big[\,1-v_u v_e\,r^4\,\delta^2\,\Big]^2}\ \ \ ,
\label{eq:Azeta0}
\end{equation}
$A_0$ being the asymmetry parameter in the SM
for arbitrary allowed $\beta$-decays,
given in (\ref{eq:Appendix.A0}) in the Appendix.
Note that in this particular case, the following identity happens
to be satisfied, independently of whether the allowed transition
is pure GT or not,
\begin{equation}
\Delta_{|_{\zeta=0}}\ =\ \frac{1}{4}\,\Bigg[
\left(\frac{A_0}{A_{|_{\zeta=0}}}\right)^2\,-\,1\,\Bigg]\ \ \ .
\end{equation}

Let us recall that these results, and in particular
the general ones in (\ref{eq:ALRSM}) and (\ref{eq:DeltaLRSM}),
are of application in the {\em most general\/}
LRSM possible, and do not assume that the parameters $\delta$
nor $(\tan\zeta)$ are small compared to the value unity. The expressions
given in (\ref{eq:ALRSM}) and (\ref{eq:DeltaLRSM})
for $A$ and $\Delta$ are {\em exact\/}, no approximation
whatsoever being implied at this stage (except for the fact that
recoil order corrections and possible charged Higgs contributions to the ratio
$R/R_0$ are neglected in the present discussion).

In order to gain some more insight into these general results, let us now
consider them in the particular case that no CP violation originates either
from the parameter $\omega$ nor from the ratio $v_{ud}$. Namely,
let us assume that the former parameter takes one of the
two values $(\omega=0)$ or $(\omega=\pi)$, and that the ratio
$v_{ud}$ is a real number, in which case
\begin{equation}
v_{ud}\,e^{i\omega}\ =\ \epsilon\,\sqrt{v_u}\ \ \ ,\ \ \
\epsilon\,=\,\pm\,1\ \ .
\end{equation}
Under these circumstances, it proves useful to define the quantities
\begin{equation}
x\ =\ \sqrt{v_u v_e}\, r^2 (\delta+t^2)\ -\
\epsilon\,\sqrt{v_e}\, r t\, (1-\delta)\ \ ,
\end{equation}
and
\begin{equation}
y\ =\ \sqrt{v_u v_e}\, r^2 (\delta+t^2)\ +\
\epsilon\,\sqrt{v_e}\, r t\, (1-\delta)\ \ ,
\end{equation}
as well as
\begin{equation}
\overline{x}\ =\ \delta t^2\ -\ \epsilon\,\sqrt{v_u}\, r t\, (1-\delta)\ \ ,
\end{equation}
and
\begin{equation}
\overline{y}\ =\ \delta t^2\ +\ \epsilon\,\sqrt{v_u}\, r t\, (1-\delta)\ \ .
\end{equation}

Indeed, one then observes that\footnote{The upper-script (CP) stands
for the fact that the expressions in the remainder of this section
are valid only when no CP violation originates either from $\omega$
nor from $v_{ud}$.}
\begin{equation}
X^{({\rm CP})}_+\ =\ y^2\ \ \ ,\ \ \
X^{({\rm CP})}_-\ =\ x^2\ \ \ ,\ \ \
Y^{({\rm CP})}\ =\ -\,x y\ \ ,
\end{equation}
as well as
\begin{equation}
Z^{({\rm CP})}_+\ =\ \left(1+\overline{y}\right)^2\ \ \ ,\ \ \
Z^{({\rm CP})}_-\ =\ \left(1+\overline{x}\right)^2\ \ \ ,\ \ \
T^{({\rm CP})}\ =\ \left(1+\overline{x}\right)\left(1+\overline{y}\right)\ \ \
{}.
\end{equation}
Therefore, under the assumptions stated above concerning $\omega$ and
$v_{ud}$, the asymmetry parameter $A$ and the quantity $\Delta$ reduce to,
respectively,
\begin{equation}
A^{({\rm CP})}\ =\ \frac{\mp\lambda^2\lambda_{J'J}
\left[(1+\overline{y})^2-y^2\right]\ -\ 2\delta_{J'J}\lambda
\sqrt{\frac{J}{J+1}}\left[(1+\overline{x})(1+\overline{y})-xy\right]}
{\left[(1+\overline{x})^2+x^2\right]\ +\ \lambda^2
\left[(1+\overline{y})^2+y^2\right]}\ \ \ ,
\end{equation}
and
\begin{displaymath}
\Delta^{({\rm CP})}\ =\ \frac{1}{2}\,
\frac{1}{\Big[\,(1+\overline{x})^2-x^2+\lambda^2
\Big((1+\overline{y})^2-y^2\Big)\Big]}\times
\end{displaymath}
\begin{displaymath}
\times\,\frac{1}{\Big[\,\mp \lambda^2\lambda_{J'J}
\Big((1+\overline{y})^2-y^2\Big)-2\delta_{J'J}\lambda\sqrt{\frac{J}{J+1}}\,
\Big((1+\overline{x})(1+\overline{y})-xy\Big)\Big]}\times
\end{displaymath}
\begin{equation}
\begin{array}{c l}
\times & \Bigg\{\,\mp\lambda^2\lambda_{J'J}\Big[\,
x^2(1+\overline{y})^2+y^2(1+\overline{x})^2+2\lambda^2 y^2
(1+\overline{y})^2\,\Big]\ -\\
&\ -\ 2\delta_{J'J}\lambda\sqrt{\frac{J}{J+1}}\,
\Big[\,x(1+\overline{y})+y(1+\overline{x})\,\Big]\,
\Big[\,x(1+\overline{x})+\lambda^2 y(1+\overline{y})\,\Big]\,\Bigg\}\ \ \ .
\end{array}
\end{equation}

In particular for allowed pure GT transitions, these results simplify to
\begin{equation}
A^{({\rm CP})}_{|_{GT}}\ =\ {A_0}_{|_{GT}}\ \frac{(1+\overline{y})^2-y^2}
{(1+\overline{y})^2+y^2}\ \ \ ,
\end{equation}
as well as
\begin{equation}
\Delta^{({\rm CP})}_{|_{GT}}\ =\ \frac{y^2(1+\overline{y})^2}
{\Big[(1+\overline{y})^2-y^2\Big]^2}\ =\ \frac{1}{4}\,\Bigg[\,
\left(\frac{{A_0}_{|_{GT}}}{A^{({\rm CP})}_{|_{GT}}}\right)^2\,-\,1\,
\Bigg]\ \ \ ,
\end{equation}
${A_0}_{|_{GT}}$ being the value of the asymmetry parameter in the SM
for allowed pure GT $\beta$-decays.

The above expressions generalise those obtained in Ref.\cite{Quin}
in the case of MLRSM for which $(r=1)$, $(v_u=1=v_e)$ and $(\omega=0)$ and
in the limit that both $\delta$ and $(t=\tan\zeta)$ are much
smaller than unity.
In contradistinction, the results derived in this
note are valid for {\em arbitrary\/}
LRSM parameters, independently of such or any other approximations
(except for recoil order corrections and possible charged Higgs contributions
to the ratio
$R/R_0$ which are ignored in the present discussion).

Nevertheless, to conclude let us consider
the limit in which both $\delta$ and
$(t=\tan\zeta)$ are indeed much smaller than unity,
still under the assumption that the parameter
$\left(v_{ud}e^{i\omega}\right)$
is real. Given the definitions,
\begin{equation}
\tilde{\delta}\ =\ \sqrt{v_u v_e}\, r^2 \delta\ \ \ ,\ \ \
\tilde{t}\ =\ \epsilon\sqrt{v_e}\, r t\ \ \ ,
\label{eq:tildedeltat}
\end{equation}
to first order in the quantities $\delta$ and $(t=\tan\zeta)$,
one then finds
\begin{equation}
x\ \simeq\ \tilde{\delta}\ -\ \tilde{t}\ \ \ ,\ \ \
y\ \simeq\ \tilde{\delta}\ +\ \tilde{t}\ \ \ ,\ \ \
\label{eq:tildexy}
\end{equation}
as well as
\begin{equation}
\overline{x}\ \simeq\ -\sqrt{\frac{v_u}{v_e}}\,\tilde{t}\ \ \ ,\ \ \
\overline{y}\ \simeq\ +\sqrt{\frac{v_u}{v_e}}\,\tilde{t}\ \ \ .
\label{eq:tildexybar}
\end{equation}
In other words, within the approximation that
$(\delta\ll 1)$ and $(\tan\zeta\ll 1)$, both $A^{({\rm CP})}$ and
$\Delta^{({\rm CP})}$
are determined by quadratic expressions in terms of the parameters
$\tilde{\delta}$ and $\tilde{t}$.
One then has,
\begin{equation}
A^{({\rm CP})}_{|_{\delta,t\ll 1}}\ \simeq\ \frac{\mp\lambda^2\lambda_{J'J}
\left[(1+\overline{y})^2-y^2\right]\ -\ 2\delta_{J'J}\lambda
\sqrt{\frac{J}{J+1}}
\left[(1+\overline{x})(1+\overline{y})-xy\right]}
{\left[(1+\overline{x})^2+x^2\right]\ +\ \lambda^2
\left[(1+\overline{y})^2+y^2\right]}\ \ \ ,
\label{eq:Alinearised}
\end{equation}
as well as
\begin{equation}
\Delta^{({\rm CP})}_{|_{\delta,t\ll 1}}\ \simeq\ \frac{1}{2}\,\Bigg\{
\frac{\mp \lambda^2\lambda_{J'J} y^2\ -\
2\delta_{J'J}\lambda\sqrt{\frac{J}{J+1}}\,xy}
{\mp \lambda^2\lambda_{J'J}\ -\
2\delta_{J'J}\lambda\sqrt{\frac{J}{J+1}}}\ +\
\frac{x^2+\lambda^2 y^2}{1+\lambda^2}\Bigg\}\ \ \ ,
\label{eq:Deltalinearised}
\end{equation}
with $x$, $y$, $\overline{x}$ and $\overline{y}$
now given in (\ref{eq:tildexy}) and (\ref{eq:tildexybar}), provided that
$(\delta\ll 1)$ and $(\tan\zeta\ll 1)$. Within these approximations
and under the assumptions that $(\omega=0)$ and that $v_{ud}$ is
real, the expression (\ref{eq:Deltalinearised})
for $\Delta^{({\rm CP})}_{|_{\delta,t\ll 1}}$
in terms of the parameters $x$ and $y$
coincides precisely with the one following from
Ref.\cite{Quin} within the context of MLRSM. In other words,
given the approximations $(\delta\ll 1)$ and $(\tan\zeta\ll 1)$
and the restriction that $\left(v_{ud}e^{i\omega}\right)$ is
real, the result obtained\cite{Quin}
for $\Delta^{({\rm CP})}_{|_{\delta,t\ll 1}}$
in the context of MLRSM remains
valid for general LRSM provided the parameters $\delta$
and $(t=\tan\zeta)$ are simply replaced by the parameters $\tilde{\delta}$
and $\tilde{t}$ defined in (\ref{eq:tildedeltat}), respectively.

The quadratic form obtained in (\ref{eq:Alinearised})
for the asymmetry parameter $A^{({\rm CP})}_{|_{\delta,t\ll 1}}$
in terms of the parameters $(v_u/v_e)$, $\tilde{\delta}$ and $\tilde{t}$
has been analysed in Ref.\cite{Oscar} already, albeit in the context
of MLRSM, namely when $(v_u=1=v_e)$,
$(\tilde{\delta}=\delta)$ and $(\tilde{t}=t)$.
For the quantity $\Delta^{({\rm CP})}_{|_{\delta,t\ll 1}}$
in (\ref{eq:Deltalinearised}),
the corresponding quadratic form is characterised by the relation
\begin{equation}
\Delta^{({\rm CP})}_{|_{\delta,t\ll 1}}\ \simeq\
\tilde{\delta}^2\ +\ 2\,\Delta_{\tilde{t}\tilde{\delta}}\,
\tilde{t}\tilde{\delta}\ +\ \Delta_{\tilde{t}\tilde{t}}\,\tilde{t}^2\ \ ,
\label{eq:quadraticDelta}
\end{equation}
with coefficients $\Delta_{\tilde{t}\tilde{\delta}}$ and
$\Delta_{\tilde{t}\tilde{t}}$ which may easily be determined
from (\ref{eq:Deltalinearised}). Therefore, any experimental
upper limit or value $\Delta_0$ for
the quantity $\Delta$ would determine an elliptic
or hyperbolic (exclusion) contour in the plane $(\tilde{t},\tilde{\delta})$,
under the conditions for which the result in (\ref{eq:Deltalinearised})
is applicable.
In particular, the slope of this contour at $(\tilde{t}=0)$,
corresponding to a vanishing mixing angle $(\zeta=0)$, is simply
given by the coefficient $(-\Delta_{\tilde{t}\tilde{\delta}})$, namely,
\begin{equation}
\left(\frac{d\tilde{\delta}}{d\tilde{t}}\right)_{|_{\zeta=0}}\ =\ -\
\Delta_{\tilde{t}\tilde{\delta}}\ \ .
\label{eq:slope1}
\end{equation}
On the other hand, the hyperbolic or elliptic character of the
contour plot is simply determined from the sign of the quantity
\begin{equation}
\Delta_{\tilde{t}\tilde{t}}\ -\
\Delta_{\tilde{t}\tilde{\delta}}^2\ \ .
\end{equation}
When this sign is positive, the contour is elliptic;
when it is negative, the contour is hyperbolic; and when this
quantity vanishes identically, the contour is a straight line.
The latter instance applies in particular to allowed pure GT decays,
since $\Delta^{({\rm CP})}_{|_{\delta,t\ll 1}}$ then reduces precisely to
$\left(y^2\simeq (\,\tilde{\delta}+\tilde{t}\,)^2\right)$.

\section{The case for mirror nuclei}
\label{Sect4}

Let us now apply the results of the previous general developments
to specific super-allowed $\beta$-decays, namely mirror nuclei.
The interest of this choice
lies with the fact that the initial and daughter nuclei
being members of a single isospin multiplet,
one may be quite confident in the determination
of recoil order corrections\cite{Holstein}
to the ratio $R/R_0$
for such nuclei using both experimental information and theoretical
considerations such as CVC and PCAC.

The list of mirror nuclei considered here\cite{Oscar}
is given in Table \ref{Table1},
with in the second column the spin sequence $(J,J')$.
Since for mirror nuclei one has $(J'=J)$, only the common value
of $J$ is indicated. Table \ref{Table1} also includes the two
allowed pure GT transitions of $^{107}$In and $^{12}$N, for
comparison. The third and fourth columns of the Table
give the values for the quantities $\lambda$ and $A_0$ introduced
previously, as evalued\footnote{The data for the neutron
are from Ref.\cite{PDG}, while a change of sign for the parameter
$\lambda$\cite{Budick} as compared to the one given in Ref.\cite{Oscar},
and the ensuing modification of the value for
the asymmetry
parameter $A_0$, are effected in the case of $^3$H.
The sign of $\lambda$ relative to that of the neutron
for all cases listed in Table~\ref{Table1} agrees with the
results given in Ref.\cite{Raman} on basis of the shell model.}
in Ref.\cite{Oscar}.
The fifth column gives the end-point total
energy $E_0$, thus including the electron (positron) rest-mass $m$,
while the sixth column lists the corresponding value $\beta_0$ of
the velocity of the $\beta^{\mp}$ particle.
The meaning of the last two columns of Table \ref{Table1} is
detailed below. Note that
except for the first two entries,
namely the neutron and $^3$H,
all nuclei listed in Table~\ref{Table1}
decay by positron emission.

In order to assess the potentiality of a given nucleus as to its
sensitivity to right-handed current contributions through a
relative longitudinal polarisation-asymmetry correlation measurement, given
the ideal experimental conditions assumed in this note---namely a
measurement of exactly the longitudinal polarisation at a specific
energy $E$ and momentum direction $\hat{p}$
of the $\beta^{\mp}$ particle---, the
following strategy is applied. The ratio $R/R_0$
is determined experimentally with a certain precision, and is
established either not to differ from the value unity by more than
that precision $\epsilon_0$, or to differ from the value unity
by a value $\epsilon_0$. In other words, in either case one may
write\footnote{In the case of the ratio $R(-J,J)/R_0(-J,J)$,
strictly speaking
the expression in (\ref{eq:limit1}) in fact ignores an additional
correction factor dependent on $\Delta$, $\beta$ and
$A_{\rm exp}(-J,J)$, given in (\ref{eq:RR02}). However, for the purpose
of the discussion of the present section, any correction brought about
by that factor may safely be ignored, since $\Delta$ is certainly
small in comparison to the value unity.}
\begin{equation}
\mid\, 1\ -\ \frac{R}{R_0}\,\mid\ =\ \mid\,k\,\Delta\,\mid\
\le\ \epsilon_0\ \ \ ,
\label{eq:limit1}
\end{equation}
where $k$ is one of the enhancement factors $k(-J,0)$ or $k(-J,J)$
depending on whether it is $R(-J,0)/R_0(-J,0)$ or $R(-J,J)/R_0(-J,J)$
which is measured. In the general case,
the quantity $\Delta$ given in (\ref{eq:DeltaLRSM})
is a rather complicated
function of the LRSM parameters $\delta$, $t$,
${\rm Re}\left(v_{ud}e^{i\omega}\right)$, $v_u$, $v_e$ and $r$, and as such
a characterisation of the potentiality of a given nucleus in terms
of the quantity $\Delta$ may not be very indicative of the new
physics it would imply. Rather, it seems more efficient to characterise
this potentiality in terms of the mass range for the $W'$ mass $M_2$
one may hope to reach with this type of experiment. For this purpose,
it is appropriate to consider the expression of the quantity $\Delta$
for a vanishing mixing angle $\zeta$, namely,
\begin{equation}
\Delta_{|_{\zeta=0}}\ =\ \frac{\tilde{\delta}^2}
{\Big[\,1-\tilde{\delta}^2\,\Big]^2}\ \ \ ,\ \ \
\tilde{\delta}^2\ =\ v_u v_e\,r^4\,\delta^2\ =\
v_u v_e\,r^4\,\frac{M_1^4}{M_2^4}\ \ ,
\end{equation}
which is valid in the most general LRSM possible (see (\ref{eq:Azeta0})\,).
Since $\tilde{\delta}^2$ is positive, one always has,
\begin{equation}
\tilde{\delta}^2\ \le\ \frac{\tilde{\delta}^2}
{\Big[\,1-\tilde{\delta}^2\,\Big]^2}\ =\ \Delta_{|{\zeta=0}}\ \ \ ,
\end{equation}
which upon substitution of the upper bound or value for
$\Delta$ given in (\ref{eq:limit1}), leads to
the following lower limit on $M_2$,
\begin{equation}
M_2\ \ge\,|r|\,\left(v_u v_e\right)^{1/4}\,M_{\rm min}\ \ \ ,
\label{eq:limit2}
\end{equation}
where the mass-reach $M_{\rm min}$ is defined as,
\begin{equation}
M_{\rm min}\ =\ \left(\frac{|k|}{\epsilon_0}\right)^{1/4}\ M_1\ \ ,
\end{equation}
$(M_1=80.22\ {\rm GeV}/c^2)$ being the mass of the ordinary gauge boson $W$.
Therefore, the potentiality of a given nucleus is to be characterised
in terms of the quantity $M_{\rm min}$, evaluated for either of the
two measurements considered here, namely corresponding
to the ratios $R(-J,0)/R_0(-J,0)$ or
$R(-J,J)/R_0(-J,J)$. Note that $M_{\rm min}$ represents precisely the lower
bound to be obtained for $M_2$ in the context of MLRSM, in the
limit that $\delta$ is much smaller than the value unity. For general
LRSM however, this is not the case and the potential lower bound on $M_2$
is obtained by multiplying the mass-reach $M_{\rm min}$ by the
factor $\Big(|r|(v_u v_e)^{1/4}\Big)$.

Clearly, the evaluation of the mass-reach $M_{\rm min}$
requires the value of the enhancement factor $k$,
which in turn needs the values for the $\beta^{\mp}$ particle velocity
$\beta$ and the experimental asymmetry $A_{\rm exp}$. The choice of value
for $\beta$ is conditioned by the necessity of high statistics measurements,
which makes it is preferable to work at the maximum of the energy distribution
$W_0(E)$. When ignoring the Coulomb correction
represented by the Fermi function $F(\pm Z,E)$, this maximum
is reached at an energy $E_{\rm max}$ whose value is given
in (\ref{eq:Emax}) of the Appendix
in terms of the end-point total energy $E_0$.
The values of $E_{\rm max}$ as well as the associated value for
the velocity $\beta_{\rm max}$ are listed in the last two columns
of Table \ref{Table1}. Thus, $\beta_{\rm max}$ is the value at which
we choose to evaluate the enhancement factors $k(-J,0)$ and $k(-J,J)$.
Accounting for Coulomb corrections through Fermi's function
$F(\pm Z,E)$ would not modify the value for $\beta_{\rm max}$
significantly, since $\beta$-decay energy spectra are typically rather
smooth around their maximum and the effect of the
shift in $\beta_{\rm max}$ due to Fermi's function is small
for small values of $Z$, as is the case for the nuclei considered
here. In any case, from the practical
experimental point of view, this issue is rather academic since
any energy acceptance is always of finite resolution.

As shown in Sect.\ref{Sect2}, the enhancement factors $k(-J,0)$
and $k(-J,J)$ are the largest when the experimental asymmetries
$A_{\rm exp}(-J,0)$ and $A_{\rm exp}(-J,J)$
approach the values $\beta^2$ and $2\beta^2/(1+\beta^2)$,
respectively.
Given the value $\beta_{\rm max}$ chosen here, these optimal
experimental asymmetries $A^{(0)}_{\rm exp}$
are listed in the last two columns of
Table \ref{Table2}.  Note that with the exception of $^3$H, these values
are rather large and thus imply the requirement of the largest
degree of nuclear polarisation possible, as is indeed to be expected.

The other data in Table \ref{Table2} give
on the one hand, the
coefficients $\Delta_{\tilde{t}\tilde{\delta}}$ and
$\Delta_{\tilde{t}\tilde{t}}$ defining the quadratic form
in (\ref{eq:quadraticDelta}) which determines the quantity
$\Delta^{({\rm CP})}_{|_{\delta,t\ll 1}}$
for values of $\delta$ and $(t=\tan\zeta)$ small compared
to the value unity and when $(v_{ud}e^{i\omega})$ is real,
and on the other hand, the type of
curve so obtained, ``E", ``H" and ``L" standing for
an elliptic, hyperbolic or linear curve, respectively.
In particular, the coefficient $(-\Delta_{\tilde{t}\tilde{\delta}})$
determines the slope $(d\tilde{\delta}/d\tilde{t})_{|_{\zeta=0}}$
at $(\zeta=0)$ of the dependence $\tilde{\delta}(\tilde{t})$
determined by $\Delta^{({\rm CP})}_{|_{\delta,t\ll 1}}$
when both $\delta$ and $(t=\tan\zeta)$
are small in comparison to the value unity;
this slope thus characterises the sensitivity
of the measurement of $\Delta$ to values of $\zeta$ different from zero
when $\delta$ is small compared to unity.
Incidentally, note that the data listed in Tables \ref{Table1}
and \ref{Table2}
are independent both of the experimental precision $\epsilon_0$ and
of the experimental asymmetries $A_{\rm exp}(-J,0)$ and
$A_{\rm exp}(-J,J)$---namely
the attainable degree of nuclear polarisation $J$---, to which
we now turn.

The experimental asymmetries
$A_{\rm exp}(-J,0)$ and $A_{\rm exp}(-J,J)$
are determined in terms of the quantity
$\Big(\beta\,J\,A\,\left(\hat{p}.\hat{J}\right)\,\Big)$.
Besides the quantity $\beta$ whose value has now been specified
to be $\beta_{\rm max}$,
one also requires the asymmetry parameter $A$ and the {\em effective\/}
degree of nuclear polarisation
$\left({\cal P}=|J\,(\hat{p}.\hat{J})|\right)$.
For the sake of the present evaluation, it is obviously justified to
approximate the former quantity by the asymmetry parameter $A_0$
in the SM, which is listed\cite{Oscar} in Table~\ref{Table1}. Indeed,
even though this approximation ignores possible right-handed contributions,
the latter are certainly small and may effectively be accounted
for through a small rescaling of the degree of nuclear
polarisation $J$.
Finally, the effective degree of nuclear polarisation is characterised
by the quantity $\Big({\cal P}=|J(\hat{p}.\hat{J})|\Big)$,
with the relative directions of $\hat{J}$ and $\hat{p}$
chosen such that the experimental
asymmetries $A_{\rm exp}(-J,0)$ and $A_{\rm exp}(-J,J)$ as defined
in this note are {\em positive\/}.

For the comparison of the potentiality offered by the nuclei considered
here, the following values for the experimental precision
$\epsilon_0$ on the measurement of $R/R_0$ and for the effective
degree of nuclear polarisation ${\cal P}$ are used,
\begin{equation}
\epsilon_0\ =\ 0.01\ \ \ ,\ \ \
{\cal P}\ =\ 0.80\ \ \ ,
\end{equation}
corresponding in fact to rather stringent experimental
requirements and achievements.
The corresponding results for the enhancement factor $k$,
the experimental asymmetry $A_{\rm exp}$ and the mass-reach
$M_{\rm min}$ are listed in Table \ref{Table3}, for both types
of configurations of nuclear polarisation
considered in this note.

The values of $M_{\min}$ in Table \ref{Table3} reveal that among
mirror nuclei, those with the best prospects with regards to our
purpose are $^{17}$F, $^{41}$Sc and $^{25}$Al, in order of
decreasing potentiality. These nuclei are also those for which
the enhancement factors are the largest, and for which the
asymmetry parameter $A_0$ is closest to the maximal value
of unity attained for allowed pure GT transitions. Indeed,
these three mirror nuclei compete well with the two
examples of the latter type of decay, namely $^{107}$In and
$^{12}$N. In fact, there are three factors---related to
one another---which concur
to explain the distinguished role of these three mirror nuclei:
a large asymmetry parameter $A_0$, allowing an experimental
asymmetry $A_{\rm exp}$ close to its optimal value $A_{\rm exp}^{(0)}$,
hence a large enhancement factor.

In spite of the large effective degree of nuclear polarisation
$({\cal P}=0.80)$ assumed here, experimental asymmetries
$A_{\rm exp}$ are still less than their optimal values
$A_{\rm exp}^{(0)}$ for all nuclei considered, given the
present choice for $\beta$, namely $(\beta=\beta_{\rm max})$.
The values for $A^{(0)}_{\rm exp}$ are quite close to unity---except
for $^3$H---simply because the values for $\beta_{\rm max}$
are also quite close to the maximal value of unity.
Indeed, as was pointed out in Sect.\ref{Sect2}, the optimal sensitivity
is achieved for values of $\beta$ and of the effective degree of
nuclear polarisation ${\cal P}$ such that $\Big(\beta=|A{\cal P}|\Big)$,
which is not possible for asymmetry parameters $|A|$ less than unity
and an effective nuclear polarisation $({\cal P}=0.80)$, once the value
of $\beta$ is specified to be $\beta_{\rm max}$.
Therefore for all nuclei considered, given this value
for $\beta$, an increased sensitivity
to right-handed charged currents requires a larger enhancement
factor, hence a larger effective degree of nuclear polarisation ${\cal P}$.
Independently of the technical feasibility of the production
and polarisation of the mirror nuclei $^{17}$F, $^{41}$Sc
and $^{25}$Al---as well as of $^{107}$In\cite{107In} and
$^{12}$N\cite{12N}---with
sufficient yields and degrees of polarisation, it thus appears that
a mass-reach $M_{\rm min}$ of the order of $600$ GeV/c$^2$
is the ultimate limit attainable
using relative longitudinal polarisation-asymmetry correlation measurements
in allowed nuclear $\beta$-decays. Note that all these cases correspond
to positron emitters, for which well established precision
polarimetry techniques are readily available\cite{107In,12N}.

However, let us remark that this conclusion which is established on quite
general grounds leaves open two possible types of loopholes. On the
one hand, there exist specific mirror nuclei for which
efficient polarisation techniques are becoming available, possibly
reaching the ideal degree of polarisation of 100\%. The above analysis
has then to be reconsidered separately for such particular cases.
On the other hand, but then at the cost of a loss of statistics,
there also remains the possibility to work at
values of $\beta$ smaller than $\beta_{\rm max}$, in order to
reach more easily the optimal sensitivity attained when $(\beta=|A{\cal P}|)$
given a certain effective degree of nuclear polarisation ${\cal P}$
achieved in practice.

The first possibility is realised for example in the cases of
$^{21}$Na\cite{Freedman} and
$^{37}$K\cite{Deutsch}. To illustrate the point,
Table~\ref{Table4} lists the same information
as Table~\ref{Table3} for a precision $(\epsilon_0=0.01)$ but
for an effective degree of nuclear
polarisation taking the maximal value possible $({\cal P}=1.00)$.
Note that $^{17}$F, $^{41}$Sc and $^{25}$Al then still remain the favorite
cases among mirror nuclei, but now with a different order of interest.
This is due to the choice of the $\beta^{\mp}$ particle energy at $E_{\rm
max}$,
which is such that for the first two cases the experimental asymmetries
$A_{\rm exp}$ are {\em larger\/} than their optimal values $A^{(0)}_{\rm exp}$.
Indeed, the enhancement factors $k(-J,0)$ and $k(-J,J)$ are now negative
for $^{17}$F and $^{41}$Sc, whereas those for $^{25}$Al remain positive
but have become large. The same applies also to the pure GT transitions
of $^{107}$In and $^{12}$N. In other words, in these specific cases,
a choice of $\beta$ slightly less than $\beta_{\rm max}$ may lead
to large enhancement factors indeed, at no significant loss
in statistics. Nevertheless, this assumes the maximal possible
effective degree of nuclear polarisation $({\cal P}=1.00)$, quite a unique
experimental circumstance. For example, even though the case of $^{12}$N
may appear from Table~\ref{Table4}
to be the most attractive with a mass-reach of
1.7~TeV/c$^2$, an effective nuclear polarisation
of $({\cal P}\simeq 0.15)$ only is obtained\cite{12N} in practice.

Therefore,
given a technically achievable effective degree of nuclear polarisation
${\cal P}$ for a specific mirror nucleus,
the other avenue open towards large enhancement factors is to consider working
at values of $\beta$ lower than $\beta_{\rm max}$.
The ensuing loss in statistics has then to be weighted
against the possibly important gain in sensitivity, but such
an evaluation is possible only on a case by case basis in contradistinction
to the general considerations of this note. However, as was remarked
previously, the $\beta$ particle count rate for the direction of
nuclear polarisation offering the greatest sensitivity to right-handed
currents is proportional to $(1-\beta^2)$ when the optimal choice
of values for $(\beta,{\cal P})$ such that $(\beta=|A{\cal P}|)$ is made.
Thus, even though the overall statistics may decrease by choosing
a value for $\beta$ smaller than $\beta_{\rm max}$ in order
to achieve a sensitivity closer to the optimal situation,
the relative statistics measured
for the direction of nuclear polarisation most sensitive to the
sought-for physical effect will increase.

One particular case which is to be distinguished from that point of view
is that of $^3$H, with a maximum value of the $\beta^-$ particle
velocity at $(\beta_0=0.2626)$, in spite of the rather small
asymmetry parameter of $(A_0=-0.09405)$. Since the corresponding value
of $(\beta_{\rm max}=0.1208)$ is quite small, working at a value of $\beta$
such that the optimal configuration $(\beta=|A{\cal P}|)$ is achieved
should not lead to any significant loss of statistics, provided
an effective degree of nuclear polarisation of at least $({\cal P}=0.80)$
can be achieved. For example, given an
effective nuclear polarisation $(0.80\le{\cal P}\le 1.00)$, the
optimal choice for $\beta$ lies in the interval
$(0.0752\le\beta\le 0.0941)$,
which is not much less than the value $(\beta_{\rm max}=0.1208)$.
Under such circumstances, given sufficient energy
resolution\footnote{A poor energy
resolution dilutes enhancement factors which otherwise could reach
very large values.}, quite large enhancement factors may be
expected, possibly opening up a mass-reach in the TeV/c$^2$ region.
Nevertheless, such instances of relative longitudinal polarisation-asymmetry
correlation measurements can be assessed on a case by case basis only.

\section{Conclusion}
\label{Sect5}

In this note, the sensitivity\cite{Quin}
of relative polarisation-asymmetry correlation $\beta$-decay experiments
to charged weak current interactions of right-handed chirality
is con\-si\-de\-red independently of any specific model for physics
beyond the Standard Model of the electroweak interactions.
Starting with the general results of Ref.\cite{JTW}
based on a four-fermi effective Hamiltonian for allowed $\beta$-decay
including arbitrary complex vector and axial coefficients
$C_V$, $C_V'$, $C_A$ and $C_A'$ only, and ignoring recoil
order corrections expected to be small for super-allowed
decays, it is shown that this class
of experiments is directly sensitive to physics beyond the SM
through a certain combination
of these four parameters which is
characterised by a single quantity $\Delta$
given in (\ref{eq:4Delta}).
A non vanishing value for $\Delta$ would establish the existence
of charged right-handed currents and thus
of new physics beyond the SM.

These general considerations are then developed further in the
specific case of so-called Left-Right Symmetric Models in their
most general form possible, assuming only that possible
charged Higgs contributions are negligible.
Which combinations of the fundamental
parameters of such LRSM are probed through the class of measurements
mentioned above is made explicit, in particular in terms
of the asymmetry parameter $A$ and the quantity $\Delta$ in
(\ref{eq:ALRSM}) and (\ref{eq:DeltaLRSM}),
respectively. These expressions, which do not
involve any simplifying restriction nor approximation whatsoever,
are also considered for restricted classes of LRSM in which no
CP violation originates from a lack of
complete complex phase alignment between
the two sectors of opposite chiralities in such theories, neither
in the Higgs nor in the Yukawa sectors.
In particular, the as\-so\-cia\-ted
results generalise those obtained previously\cite{Quin}
in the context of so-called Manifest Left-Right Symmetric Models---which
provide but one type of a very restricted class of LRSM---under
the approximation that both the ratio of
the squared masses of light to heavy charged
gauge bosons $W$ and $W'$ as well as their mixing angle be much
smaller than the value unity.

These general results are then applied specifically to the case
of mirror nuclei, which offer the advantage that recoil order corrections
are more amenable to sufficiently precise evaluation than for
other instances of allowed $\beta$-decay, since initial
and daughter nuclei then belong to the same isospin multiplet.
The potentiality of these mirror nuclei as to the sensitivity
to contributions from charged currents of right-handed chirality
is then characterised in terms of the mass-reach---the
precise technical meaning of this notion in the context
of general LRSM is defined in Sect.\ref{Sect4}---for
the hypothetical $W'$ charged gauge boson which may be achievable
by using each of these nuclei,
given a certain experimental precision and
degree of nuclear polarisation. The analysis es\-ta\-bli\-shes that among
mirror decays, the cases of $^{17}$F, $^{41}$Sc and $^{25}$Al,
in order of decreasing interest,
certainly offer the best prospects, which are comparable to those
achieved by on-going
experiments using the allowed pure Gamow-Teller transitions of
$^{107}$In\cite{107In} and $^{12}$N\cite{12N}.
Indeed, allowed pure Gamow-Teller decays are expected
to provide the best mass-reach possible owing to their
maximal asymmetry parameter.

The analysis is performed on quite general grounds, not paying
attention to specific circumstances which may apply
to a given particular nucleus, nor to the technical feasibility
of the production and polarisation of these nuclei.
In fact, the sensitivity to right-handed
charged current contributions of the type of experiment considered
here may become quite large for some special cases, by appropriatedly
choosing to work at a specific value of the
$\beta^\mp$ particle energy, given an achievable effective
degree of nuclear polarisation. It is then not excluded that
some particular mirror nucleus presents the potential to extend
the mass-reach of relative longitudinal polarisation-asymmetry
correlation measurements into the TeV/c$^2$ region. One such
instance which may be worth pursuing further could be that of
the $\beta^-$-decay of $^3$H, owing to the rather low value
of the end-point energy in that case.

This conclusion also opens the prospect that for specific
values of the $\beta^\mp$ particle energy and of the nuclear polarisation,
other observables
in the $\beta$-decay of mirror nuclei
offer a similarly large sensitivity to other
couplings appearing\cite{JTW} in the effective Hamiltonian
for nuclear $\beta$-decay, including for example scalar
or tensor contributions, as well as time reversal violating
effects\cite{Deutsch}. Such possibilities certainly deserve to be
investigated in detail, along lines similar to those developed here.

\section*{Acknowledgement}

Drs. O. Naviliat-Cuncic and
N. Severijns are gratefully acknowledged for useful information
concerning the value of $\lambda$
in the case of $^3$H.

\pagebreak

\section*{Appendix}
\label{Appendix}

In the case of vector ($V$) and axial ($A$) contributions only,
as assumed in this note, the
general four-fermi effective Hamiltonian considered in Ref.\cite{JTW}
at the nucleon level is of the form
\begin{equation}
H^{\rm nucleon}_{\rm eff}\ =\ \overline{\psi}_{p}\gamma_\mu\psi_n\,
\overline{\psi}_e\,\Big(C_V\gamma^\mu-{C_V}'
\gamma^\mu\gamma_5\Big)\psi_{\nu_e}
\ -\ \overline{\psi}_p\gamma_\mu\gamma_5\psi_n\,
\overline{\psi}_e\,\Big(C_A\gamma^\mu\gamma_5-{C_A}'\gamma^\mu\Big)\psi_{\nu_e}
\ \ ,
\label{eq:Heff1}
\end{equation}
where $C_V$, $C_V'$, $C_A$ and $C_A'$ are {\em a priori\/} arbitrary complex
coefficients, and $\psi_p$, $\psi_n$, $\psi_e$ and $\psi_{\nu_e}$
represent Dirac spinors for the proton, the neutron,
the electron and the neutrino
of electronic flavour, respectively. Our phase conventions are as follows.
The chirality operator $\gamma_5$ is defined with a sign such that
left-handed couplings are of the form $\gamma_\mu(1-\gamma_5)$,
which is the choice opposite to that taken in Ref.\cite{JTW}.
To account for that difference, changes of sign have been included in
the expression in (\ref{eq:Heff1}) in such a way that the coefficients
$C_V$, $C_V'$, $C_A$ and $C_A'$ are as defined in Ref.\cite{JTW}.

The quantities $W_0(E)$, $\xi$, $A$, $G$ and $N'$ appearing in
(\ref{eq:Dist1}) are then given by the following expressions\cite{JTW},
\vspace{10pt}
\begin{equation}
\xi=M_F^2\Big[\,|C_V|^2+|C_V'|^2\,\Big]\ +\
M_{GT}^2\Big[\,|C_A|^2+|C_A'|^2\,\Big]\ \ ,
\end{equation}
\vspace{10pt}
\begin{equation}
\begin{array}{r c l}
\xi A&=&M_{GT}^2\,\lambda_{J'J}\Big[\,\mp 2\,{\rm Re}
\left(C_A {C_A'}^*\right)\,\Big]\ +\ \\ \\
&+& \delta_{J'J}M_F M_{GT}\,\sqrt{\frac{J}{J+1}}\,
\Big[\,-2\,{\rm Re}\left(C_V{C_A'}^*+C_V'{C_A}^*\right)\,\Big]\ \ ,
\end{array}
\end{equation}
\vspace{10pt}
\begin{equation}
\xi G=M_F^2\,\Big[\,\mp\,2\,{\rm Re}
\left(C_V{C_V'}^*\right)\,\Big]\ +\ M_{GT}^2\,
\Big[\,\mp\,2\,{\rm Re}\left(C_A{C_A'}^*\right)\,\Big]\ \ ,
\end{equation}
\vspace{10pt}
\begin{equation}
\begin{array}{r c l}
\xi N'&=&M_{GT}^2\,\lambda_{J'J}\,\Big[\,
|C_A|^2+|C_A'|^2\,\Big]\ +\ \\ \\
&+& 2\delta_{J'J}\,M_F M_{GT}\,
\sqrt{\frac{J}{J+1}}\,\Big[\,\pm{\rm Re}
\left(C_V{C_A}^*+C_V'{C_A'}^*\right)\,\Big]\ \ ,
\end{array}
\end{equation}
\vspace{10pt}
and finally
\begin{equation}
W_0(E)\ =\ \frac{1}{(2\pi)^4}\,p\,E(E_0-E)^2\,F(\pm Z,E)\ \ .
\end{equation}
Here, $E_0$ is the $\beta$-spectrum end-point total energy, $p$ and $E$
the momentum and total energy of the $\beta^{\mp}$ particle,
respectively, and $F(\pm Z,E)$
Fermi's function for Coulomb corrections, $Z$ being the charge of
the {\em daughter\/} nucleus. For an allowed transition from an initial
state of nuclear spin $J$ to a final state of nuclear spin $J'$, the quantity
$\lambda_{J'J}$ is defined by\cite{JTW}
\begin{equation}
\lambda_{J'J}\ =\ \left\{\begin{array}{c l}
			             1 & J\rightarrow J'=J-1 \\ \\
			 \frac{1}{J+1} & J\rightarrow J'=J \\ \\
			-\frac{J}{J+1} & J\rightarrow J'=J+1
			  \end{array}\right.\ \ .
\end{equation}
And finally, $M_F$ and $M_{GT}$ are the Fermi and Gamow-Teller nucleon matrix
elements, respectively.

Ignoring the factor $F(\pm Z,E)$, it is possible to show that
the function $W_0(E)$ reaches its maximal value for a total energy
$E_{\rm max}$ given by
\begin{equation}
\frac{E_{\rm max}}{E_0}\ =\ \frac{1}{6}\ +\ \rho_E\,
\sin\Bigg\{\,\frac{1}{3}\arcsin
\Bigg[\frac{1}{\rho_E^3}
\left(\frac{1}{2}\left(\frac{m}{E_0}\right)^2-\frac{1}{27}\right)\Bigg]\
+\ \frac{2\pi}{3}\,\Bigg\}\ \ \ ,
\label{eq:Emax}
\end{equation}
where
\begin{equation}
\rho_E\ =\ \sqrt{\left(\frac{m}{E_0}\right)^2\,+\,\frac{1}{9}}\ \ \ ,
\end{equation}
$m$ being of course the electron (positron) mass.

Note that in terms of the quantities $a_L$, $a_R$, $b_L$ and
$b_R$ introduced in (\ref{eq:aL}) to (\ref{eq:bR}) of Sect.\ref{Sect2},
one may also write
\begin{equation}
\xi\ =\ \frac{1}{2}\,\left(a_L\,+\,a_R\right)\ \ \ ,
\end{equation}
\begin{equation}
\xi A\ =\ \frac{1}{2}\,\left(b_L\,-\,b_R\right)\ \ \ ,
\end{equation}
\begin{equation}
\xi G\ =\ \mp\,\frac{1}{2}\,\left(a_L\,-\,a_R\right)\ \ \ ,
\end{equation}
\begin{equation}
\xi N'\ =\ \mp\,\frac{1}{2}\,\left(b_L\,+\,b_R\right)\ \ \ .
\end{equation}

Moreover, in the particular case of allowed pure GT transitions,
one observes that
\begin{equation}
(\xi G)_{|_{GT}}\ =\ \frac{1}{\lambda_{J'J}}\,(\xi A)_{|_{GT}}\ \ \ ,\ \ \
(\xi N')_{|_{GT}}\ =\ \lambda_{J'J}\,\xi_{|_{GT}}\ \ \ ,
\end{equation}
thus showing that in this instance only the matrix element $\xi_{|_{GT}}$
and the asymmetry parameter $A_{|_{GT}}$ are relevant to the description
of the decay. In particular, relative measurements as those considered
in this note are then only dependent on the asymmetry parameter $A_{|_{GT}}$
in the case of allowed pure GT transitions.

In the Standard Model, the coefficients
$C_V$, $C_V'$, $C_A$ and $C_A'$ are simply
determined up to a common factor $C_V^0$ by the relations,
\begin{equation}
C_V'\ =\ C_V\ \ \ ,\ \ \
C_A'\ =\ C_A\ \ \ ,
\end{equation}
together with
\begin{equation}
C_V\ =\ C^0_V\ \ \ ,\ \ \
C_A\ =\ \frac{g_A}{g_V}\,C^0_V\ \ \ ,
\end{equation}
$g_V$ and $g_A$ being the standard vector and axial couplings of
nucleon $\beta$-decay, respectively, such that\cite{PDG}
\begin{equation}
\frac{g_A}{g_V}\ =\ -1.2573\pm 0.0028\ \ .
\end{equation}
Given the definitions
\begin{equation}
\rho\ =\ \frac{g_A}{g_V}\,<\,0\ \ ,\ \ \
\lambda\ =\ \frac{g_A}{g_V}\,\frac{M_{GT}}{M_F}\ =\
\rho\, \frac{M_{GT}}{M_F}\ \ \ ,
\end{equation}
one then obtains,
\begin{equation}
\xi_0=2\,|C^0_V|^2\,\Big[\,M_F^2+\rho^2 M_{GT}^2\,\Big]=
2\,|C^0_V|^2\,M_F^2\,\Big[\,1+\lambda^2\,\Big]\ \ \ ,
\end{equation}
\vspace{10pt}
\begin{equation}
\begin{array}{c c l}
(\xi A)_0&=&2\,|C^0_V|^2\,\Big[\,\mp\rho^2 M_{GT}^2\lambda_{J'J}\ -\
2\,\delta_{J'J}\,\rho\,M_F M_{GT}\,\sqrt{\frac{J}{J+1}}\,\Big]\\ \\
   & = & 2\,|C^0_V|^2\,M_F^2\,\Big[\,\mp\lambda^2\lambda_{J'J}\ -\
2\,\delta_{J'J}\lambda\,\sqrt{\frac{J}{J+1}}\,\Big]\ \ ,
\end{array}
\end{equation}
\vspace{10pt}
\begin{equation}
(\xi G)_0=\mp\,\xi_0\ \ \ ,
\end{equation}
and finally
\begin{equation}
(\xi N')_0=\mp\,(\xi A)_0\ \ \ .
\end{equation}
Thus for example, the asymmetry parameter $A_0$ in the SM
simply reduces to
\begin{equation}
A_0\ =\ \frac{1}{1+\lambda^2}\,
\Big[\,\mp\lambda^2\lambda_{J'J}\ -\
2\,\delta_{J'J}\lambda\,\sqrt{\frac{J}{J+1}}\,\Big]\ \ .
\label{eq:Appendix.A0}
\end{equation}
In particular,
for allowed pure GT transitions this result becomes
\begin{equation}
{A_0}_{|_{GT}}\ =\ \mp\lambda_{J'J}\ \ \ ,
\label{eq:Appendix.A0GT}
\end{equation}
which is thus maximal only for transitions such that $(J'=J-1)$,
\begin{equation}
{A_0}_{|_{GT}}\ =\ \mp\, 1\ \ \ ,\ \ \ J'=J-1\ \ .
\end{equation}
This is the case for example for $^{107}$In and $^{12}$N.

In the case of LRSM, the coefficients $C_V$, $C_V'$, $C_A$
and $C_A'$ are given in (\ref{eq:coeffLRSM}) in terms of the
combinations of fundamental parameters of LRSM defined
in (\ref{eq:comb1}) to (\ref{eq:comb4}). In order to list
the expressions required for the evaluation of the
parameters $\xi$, $A$, $G$ and $N'$, let us also introduce
the notation
\begin{equation}
C^2_N\ =\ 2\,g_V^2\,\eta_0^2|V^L_{ud}|^2\,{\sum_i}'|U^L_{ie}|^2\ \ \ ,
\end{equation}
and refer to the relations (\ref{eq:Xplus}) to
(\ref{eq:T}) in Sect.\ref{Sect3}
for the definition of the other quantities appearing in
the expressions which follow.
One then obtains,
\begin{equation}
|C_V|^2+|C_V'|^2=C^2_N\Bigg[\,Z_-\,+\,X_-\,\Bigg]\ \ \ ,
\end{equation}
\begin{equation}
|C_A|^2+|C_A'|^2=\rho^2\,C^2_N\Bigg[\,Z_+\,+\,X_+\,\Bigg]\ \ \ ,
\end{equation}
\begin{equation}
2\,{\rm Re}\Bigg(C_V {C_V'}^*\Bigg)=C^2_N\Bigg[\,Z_-\,-\,X_-\,\Bigg]\ \ \ ,
\end{equation}
\begin{equation}
2\,{\rm Re}\Bigg(C_A {C_A'}^*\Bigg)=\rho^2\,
C^2_N\Bigg[\,Z_+\,-\,X_+\,\Bigg]\ \ \ ,
\end{equation}
\begin{equation}
{\rm Re}\Bigg(C_V {C_A'}^*+C_V'{C_A}^*\Bigg)\,=\,\rho\,
C^2_N\Bigg[\,T\,+\,Y\,\Bigg]\ \ \ ,
\end{equation}
\begin{equation}
{\rm Re}\Bigg(C_V C_A^*+C_V'{C_A'}^*\Bigg)\,=\,\rho\,
C^2_N\Bigg[\,T\,-\,Y\,\Bigg]\ \ \ ,
\end{equation}

\vspace{10pt}

\begin{equation}
|C_V+C_V'|^2=2\,C^2_N\,Z_-\ \ \ ,
\end{equation}

\begin{equation}
|C_V-C_V'|^2=2\,C^2_N\,X_-\ \ \ ,
\end{equation}

\begin{equation}
|C_A+C_A'|^2=2\,\rho^2\,C^2_N\,Z_+\ \ \ ,
\end{equation}

\begin{equation}
|C_A-C_A'|^2=2\,\rho^2\,C^2_N\,X_+\ \ \ ,
\end{equation}

\begin{equation}
{\rm Re}\Bigg(C_V+C_V'\Bigg)\Bigg({C_A}^*+{C_A'}^*\Bigg)=2\,\rho\,
C^2_N\,T\ \ \ ,
\end{equation}

\begin{equation}
{\rm Re}\Bigg(C_V-C_V'\Bigg)\Bigg({C_A}^*-{C_A'}^*\Bigg)=-\,2\,\rho\,
C^2_N\,Y\ \ \ .
\end{equation}

\pagebreak

\begin{table}
\begin{center}
\begin{tabular}{||c||c|c c|c c|c c||}
\hline\hline
	  &     &	  &	    &	    &	      &	       &	\\
Isotope & $(J,J')$ & $\lambda$ & $A_0$ & $E_0$ & $\beta_0$ &
$E_{\rm max}$   & $\beta_{\rm max}$ \\
          &     &         &         & (MeV) &         & (MeV)  &        \\
\hline\hline
	  &     &	  &	    &	    &	      &	       &	\\
$n$       & 1/2 & -2.178  & -0.1127 & 1.293 & 0.9186 & 0.7580 & 0.7386 \\
$^{3}$H   & 1/2 & -2.095  & -0.09405& 0.5296& 0.2626 & 0.5148 & 0.1208 \\
$^{11}$C  & 3/2 &  0.7390 & -0.5992 & 1.472 & 0.9378 & 0.8326 & 0.7895 \\
$^{13}$N  & 1/2 &  0.5552 & -0.3330 & 1.701 & 0.9538 & 0.9328 & 0.8366 \\
$^{15}$O  & 1/2 & -0.6272 &  0.7080 & 2.241 & 0.9737 & 1.1815 & 0.9016 \\
$^{17}$F  & 5/2 & -1.2860 &  0.9972 & 2.259 & 0.9741 & 1.1899 & 0.9031 \\
$^{19}$Ne & 1/2 &  1.6000 & -0.0396 & 2.751 & 0.9826 & 1.4245 & 0.9334 \\
$^{21}$Na & 3/2 & -0.7039 &  0.8617 & 3.021 & 0.9856 & 1.5549 & 0.9445 \\
$^{23}$Mg & 3/2 &  0.5400 & -0.5574 & 3.610 & 0.9899 & 1.8419 & 0.9607 \\
$^{25}$Al & 5/2 & -0.8004 &  0.9362 & 3.781 & 0.9908 & 1.9256 & 0.9641 \\
$^{27}$Si & 5/2 &  0.6870 & -0.6973 & 4.361 & 0.9931 & 2.2108 & 0.9729 \\
$^{29}$P  & 1/2 & -0.5112 &  0.6060 & 4.456 & 0.9934 & 2.2577 & 0.9740 \\
$^{31}$S  & 1/2 &  0.5131 & -0.3301 & 4.901 & 0.9945 & 2.4774 & 0.9785 \\
$^{33}$Cl & 3/2 &  0.2888 & -0.3821 & 5.021 & 0.9948 & 2.5368 & 0.9795 \\
$^{35}$Ar & 3/2 & -0.2722 &  0.4201 & 5.451 & 0.9956 & 2.7497 & 0.9826 \\
$^{37}$K  & 3/2 &  0.5811 & -0.5720 & 5.641 & 0.9959 & 2.8438 & 0.9837 \\
$^{39}$Ca & 3/2 & -0.6424 &  0.8213 & 6.001 & 0.9964 & 3.0224 & 0.9856 \\
$^{41}$Sc & 7/2 & -1.0664 &  0.9983 & 6.121 & 0.9965 & 3.0820 & 0.9862 \\
	  &     &	  &	    &	    &	      &	       &	\\
$^{12}$N  &(1,0)&$\infty$ &  1.0000 & 16.891& 0.9995 & 8.4532 & 0.9982 \\
$^{107}$In&(9/2,7/2)&$\infty$&1.0000& 2.761 & 0.9827 & 1.4293 & 0.9339 \\
	  &     &	  &	    &	    &	      &	       &	\\
\hline\hline
\end{tabular}
\caption[]{List of mirror nuclei considered in Sect.\ref{Sect4} and their
characteristics, including the two allowed pure Gamow-Teller
decays of $^{107}$In\cite{107In} and $^{12}$N\cite{12N} for
comparison. The last column gives the value of $\beta$ at
which the potentiality of each nucleus is evaluated. Further details
are given in the text.}
\label{Table1}
\end{center}
\end{table}

\pagebreak

\begin{table}
\begin{center}
\begin{tabular}{||c||c c c|c c c||}
\hline\hline
	  &          &	        &   & &	      &         \\
Isotope & $\Delta_{\tilde{t}\tilde{\delta}}$ & $\Delta_{\tilde{t}\tilde{t}}$ &
Type & & $\beta_{\rm max}^2$ &
$2\,\beta_{\rm max}^2/(1+\beta_{\rm max}^2)$ \\
	  &          &	        &   & &	      &	        \\
\hline\hline
	  &          &	        &   & &	      &	        \\
$n$       &  2.769   &  4.887   & H & & 0.5455  & 0.7059  \\
$^{3}$H   &  3.200  &   5.771   & H & & 0.0146  & 0.0288  \\
$^{11}$C  & -0.2647  & -0.2358  & H & & 0.6233  & 0.7680  \\
$^{13}$N  & -0.5003  & -0.4718  & H & & 0.6999  & 0.8235  \\
$^{15}$O  & -0.08476 &  0.2658  & E & & 0.8129  & 0.8968  \\
$^{17}$F  &  0.2125  &  0.1786  & E & & 0.8156  & 0.8984  \\
$^{19}$Ne & -5.839   & -12.12   & H & & 0.8713  & 0.9312  \\
$^{21}$Na & -0.09177 &  0.1538  & E & & 0.8920  & 0.9429  \\
$^{23}$Mg & -0.3553  & -0.1620  & H & & 0.9230  & 0.9600  \\
$^{25}$Al & -0.0499  &  0.1192  & E & & 0.9296  & 0.9635  \\
$^{27}$Si & -0.2451  & -0.1314  & H & & 0.9466  & 0.9726  \\
$^{29}$P  & -0.1788  &  0.2279  & E & & 0.9488  & 0.9737  \\
$^{31}$S  & -0.5021  & -0.4209  & H & & 0.9575  & 0.9783  \\
$^{33}$Cl & -0.4633  & -0.08057 & H & & 0.9594  & 0.9793  \\
$^{35}$Ar & -0.3982  &  0.06566 & H & & 0.9655  & 0.9824  \\
$^{37}$K  & -0.3358  & -0.1765  & H & & 0.9677  & 0.9836  \\
$^{39}$Ca & -0.1368  &  0.1423  & E & & 0.9714  & 0.9855  \\
$^{41}$Sc &  0.09134 &  0.1184  & E & & 0.9725  & 0.9861  \\
	  &          &	        &   & &	      &	        \\
$^{12}$N  &  1.000   &  1.000   & L & & 0.9963  & 0.9982  \\
$^{107}$In&  1.000   &  1.000   & L & & 0.8722  & 0.9317  \\
	  &          &	        &   & &	      &	        \\
\hline\hline
\end{tabular}
\caption[]{Coefficients of the quadratic form in
(\ref{eq:quadraticDelta}) characterising the parameter $\Delta$
for values of $\delta$ and $(\tan\zeta)$ much smaller
than unity and for $(v_{ud}e^{i\omega})$ real. The symbols
``E", ``H" or ``L" stand for whether this quadratic form
determines an elliptic, hyperbolic or linear curve
in the plane $(\tilde{t},\tilde{\delta})$, respectively
(further details are given in the text). The last
two columns give the optimal values $A_{\rm exp}^{(0)}$
for the experimental
asymmetries $A_{\rm exp}(-J,0)$ and $A_{\rm exp}(-J,J)$,
respectively, when $(\beta=\beta_{\rm max})$.}
\label{Table2}
\end{center}
\end{table}

\pagebreak

\begin{table}
\begin{center}
\begin{tabular}{||c||c c c||c c c||}
\hline\hline
	  &         &	     &	   &	    &        &      \\
Isotope & $k(-J,0)$ & $A_{\rm exp}(-J,0)$ & $M_{\rm min}$ &
$k(-J,J)$ & $A_{\rm exp}(-J,J)$ & $M_{\rm min}$ \\
	  &         &	     & (GeV/c$^2$) &    & & (GeV/c$^2$) \\
\hline\hline
	  &         &	     &	   &	    &        &      \\
$n$       &  0.556  & 0.0666 & 219 &  0.991 & 0.125  & 253  \\
$^{3}$H   &  6.60   & 0.0091 & 407 &  8.14  & 0.018  & 428  \\
$^{11}$C  &  6.18   & 0.378  & 400 &  7.69  & 0.549  & 422  \\
$^{13}$N  &  1.87   & 0.223  & 297 &  2.83  & 0.365  & 329  \\
$^{15}$O  &  6.76   & 0.511  & 409 &  8.30  & 0.676  & 431  \\
$^{17}$F  & 30.3    & 0.720  & 595 &  32.2  & 0.838  & 604  \\
$^{19}$Ne &  0.141  & 0.0296 & 155 &  0.272 & 0.0574 & 183  \\
$^{21}$Na & 10.8    & 0.651  & 460 & 12.5   & 0.789  & 477  \\
$^{23}$Mg &  3.46   & 0.428  & 346 &  4.73  & 0.600  & 374  \\
$^{25}$Al & 13.9    & 0.722  & 490 & 15.7   & 0.839  & 505  \\
$^{27}$Si &  5.38   & 0.543  & 386 &  6.83  & 0.704  & 410  \\
$^{29}$P  &  3.96   & 0.472  & 358 &  5.29  & 0.642  & 385  \\
$^{31}$S  &  1.48   & 0.258  & 280 &  2.33  & 0.411  & 313  \\
$^{33}$Cl &  1.81   & 0.299  & 294 &  2.77  & 0.461  & 327  \\
$^{35}$Ar &  2.08   & 0.330  & 305 &  3.10  & 0.496  & 337  \\
$^{37}$K  &  3.48   & 0.450  & 346 &  4.75  & 0.621  & 374  \\
$^{39}$Ca &  8.00   & 0.648  & 427 &  9.60  & 0.786  & 447  \\
$^{41}$Sc & 17.0    & 0.788  & 515 & 18.8   & 0.881  & 528  \\
	  &         &	     &	   &	    &        & 	    \\
$^{12}$N  & 16.2    & 0.799  & 509 & 17.9   & 0.888  & 522  \\
$^{107}$In& 23.9    & 0.747  & 561 & 25.7   & 0.855  & 571  \\
	  &         &	     &	   &	    &	     &      \\
\hline\hline
\end{tabular}
\caption[]{The enhancement factors $k$, experimental asymmetries
$A_{\rm exp}$
and mass-reaches $M_{\rm min}$
for $(\epsilon_0=0.01)$, $({\cal P}=0.80)$ and $(\beta=\beta_{\rm max})$.}
\label{Table3}
\end{center}
\end{table}

\clearpage

\pagebreak

\begin{table}
\begin{center}
\begin{tabular}{||c||c c c||c c c||}
\hline\hline
	  &         &	     &	   &	    &        &      \\
Isotope & $k(-J,0)$ & $A_{\rm exp}(-J,0)$ & $M_{\rm min}$ &
$k(-J,J)$ & $A_{\rm exp}(-J,J)$ & $M_{\rm min}$ \\
	  &         &	     & (GeV/c$^2$) &    & & (GeV/c$^2$) \\
\hline\hline
	  &         &	     &	   &	    &        &      \\
$n$       &  0.720  & 0.0832 & 234 &  1.25  & 0.154  & 268  \\
$^{3}$H   & 14.1    & 0.0114 & 491 & 15.8   & 0.0225 & 506  \\
$^{11}$C  & 12.6    & 0.473  & 478 & 14.3   & 0.642  & 493  \\
$^{13}$N  &  2.64   & 0.279  & 324 &  3.78  & 0.436  & 354  \\
$^{15}$O  & 14.6    & 0.638  & 496 & 16.4   & 0.779  & 510  \\
$^{17}$F  &-42.4    & 0.901  & 647 & -40.3  & 0.948  & 639  \\
$^{19}$Ne &  0.177  & 0.0370 & 165 &  0.340 & 0.0713 & 194  \\
$^{21}$Na & 41.7    & 0.814  & 644 & 43.6   & 0.897  & 652  \\
$^{23}$Mg &  5.53   & 0.536  & 389 &  7.00  & 0.698  & 413  \\
$^{25}$Al &134.0    & 0.903  & 863 &136.0   & 0.949  & 866  \\
$^{27}$Si & 10.1    & 0.678  & 452 & 11.8   & 0.808  & 470  \\
$^{29}$P  &  6.59   & 0.590  & 406 &  8.12  & 0.742  & 428  \\
$^{31}$S  &  2.04   & 0.323  & 303 &  3.05  & 0.488  & 335  \\
$^{33}$Cl &  2.56   & 0.374  & 321 &  3.68  & 0.545  & 351  \\
$^{35}$Ar &  2.99   & 0.413  & 334 &  4.19  & 0.584  & 363  \\
$^{37}$K  &  5.56   & 0.563  & 389 &  7.03  & 0.720  & 413  \\
$^{39}$Ca & 20.0    & 0.809  & 536 & 21.8   & 0.895  & 548  \\
$^{41}$Sc &-329     & 0.984  &1080 &-327    & 0.992  &1079  \\
	  &         &	     &	   &	    &        & 	    \\
$^{12}$N  &-2187    & 0.998  &1735 &-2185   & 0.999  &1734  \\
$^{107}$In&-60.5    & 0.934  & 708 &-58.5   & 0.966  & 701  \\
	  &         &	     &	   &	    &	     &      \\
\hline\hline
\end{tabular}
\caption[]{The enhancement factors $k$, experimental asymmetries
$A_{\rm exp}$
and mass-reaches $M_{\rm min}$
for $(\epsilon_0=0.01)$, $({\cal P}=1.00)$ and $(\beta=\beta_{\rm max})$.}
\label{Table4}
\end{center}
\end{table}

\end{document}